\RequirePackage{fix-cm}

\documentclass[smallextended]{svjour3}
\smartqed

\usepackage{graphicx}
 \usepackage{mathptmx}      
\usepackage{hyperref}
\usepackage{amsmath}
\usepackage{cite}
\usepackage{bm}
\usepackage{amssymb}

\numberwithin{equation}{section}

\begin{document}

\title{Vacuum polarization on three-dimensional anti-de Sitter space-time with Robin boundary conditions
}

\titlerunning{VP on global adS${}_{3}$ with Robin boundary conditions}        

\author{Sivakumar Namasivayam \and Elizabeth Winstanley 
}

\institute{Consortium for Fundamental Physics, School of Mathematics and Statistics,  The University of Sheffield, Hicks Building, Hounsfield Road, Sheffield. S3 7RH United Kingdom
	\\	
	\email{SNamasivayam1@sheffield.ac.uk, E.Winstanley@sheffield.ac.uk}
}

\date{}

\maketitle

\begin{abstract}
We study a quantum scalar field, with general mass and coupling to the scalar curvature, propagating on three-dimensional global anti-de Sitter space-time. We determine the vacuum and thermal expectation values of the square of the field, also known as the vacuum polarisation (VP). 
We consider values of the scalar field mass and coupling for which there is a choice of boundary conditions giving well-posed classical dynamics. 
We apply Dirichlet, Neumann and Robin (mixed) boundary conditions to the field at the space-time boundary. We find finite values of the VP when the parameter governing the Robin boundary conditions is below a certain critical value.
For all couplings, the vacuum expectation values of the VP with either Neumann or Dirichlet boundary conditions are constant and respect the maximal symmetry of the background space-time. 
However, this is not the case for Robin boundary conditions, when both the vacuum and thermal expectation values depend on the space-time location. 
At the space-time boundary, we find that both the vacuum and thermal expectation values of the VP with Robin boundary conditions converge to the result when Neumann boundary conditions are applied, except in the case of Dirichlet boundary conditions. 
\end{abstract}

\section{Introduction}
\label{sec:intro}

Quantum field theory (QFT) on a background anti-de Sitter space-time (adS) has attracted much attention in recent years, not least because  of its role in the adS/CFT correspondence (see \cite{Ahorony;2000} for a review). 
While the maximal symmetry of adS simplifies many aspects of QFT on this geometry, 
the presence of a time-like boundary at spatial infinity means that adS is not a globally hyperbolic space-time. Thus, we need  to impose appropriate boundary conditions  on the time-like boundary in order to have a well-posed QFT~\cite{Avis:1977yn,Barroso:2019cwp,Dappiaggi:2017wvj,Dappiaggi:2018xvw,Dappiaggi:2018pju,Ishibashi:2004wx,Benini:2017dfw,Ishibashi:2003jd,Wald:1980jn,Dappiaggi:2021wtr,Gannot:2018jkg}.

The simplest boundary conditions to consider are Dirichlet, where the value of the field vanishes at the boundary \cite{Allen,Ambrus:2018olh,Avis:1977yn}  and Neumann, where the normal derivative of the field vanishes at the boundary \cite{Allen,Avis:1977yn,Morley:2021}. 
For these boundary conditions,  the vacuum Green's functions and vacuum expectation values (v.e.v.s) respect the maximal symmetry of the background adS~\cite{Avis:1977yn, Kent:2014nya}. 
This means that the v.e.v.~of the square of a quantum scalar field (hereafter termed the vacuum polarization (VP)) is a constant, while the v.e.v.~of the stress-energy tensor (SET) for any quantum field is a constant multiplied by the space-time metric.

Other boundary conditions can also result in well-defined field dynamics \cite{Barroso:2019cwp,Dappiaggi:2017wvj,Dappiaggi:2018xvw,Dappiaggi:2018pju,Ishibashi:2004wx,Benini:2017dfw,Ishibashi:2003jd,Wald:1980jn,Dappiaggi:2021wtr,Gannot:2018jkg}. 
For example, one can consider a linear combination of the Dirichlet and Neumann boundary conditions (known as Robin or mixed boundary conditions~\cite{Ishibashi:2004wx,Dappiaggi:2018xvw}).
Such Robin boundary conditions are the focus of this paper, however it should be emphasised that these are not the only possibilities (see, for example, \cite{Dappiaggi:2017wvj,Dappiaggi:2018pju,Dappiaggi:2021wtr} for discussions of more general boundary conditions).

The effect of Robin boundary conditions on renormalized expectation values was studied in \cite{Barroso:2019cwp,Morley:2021} for a massless, conformally coupled scalar field on global four-dimensional adS. 
Applying Robin boundary conditions to just the $s$-wave modes \cite{Barroso:2019cwp}, it is found that the v.e.v. of the VP is no longer a constant, and the v.e.v.~of the SET
is also not maximally symmetric. 
At the space-time boundary, in this scenario the v.e.v.s of both the VP and the SET approach those when Dirichlet boundary conditions are applied \cite{Barroso:2019cwp}.
Somewhat different results for the VP are obtained if Robin boundary conditions are applied to {\em {all}} field modes~\cite{Morley:2021}.
In particular, at the space-time boundary it is found that the renormalized VP for all boundary conditions approaches the value obtained when Neumann boundary conditions are applied, except in the case of Dirichlet boundary conditions~\cite{Morley:2021}.

The work of~\cite{Barroso:2019cwp,Morley:2021} considers only a massless, conformally coupled scalar field. 
What happens for more general scalar field mass and coupling?  
The maximal symmetry of the v.e.v.s of the VP and SET is preserved for arbitrary mass and coupling when Dirichlet boundary conditions are applied \cite{Kent:2014nya}, for any number of space-time dimensions.
While the ground state Green's function for a quantum scalar field with arbitrary mass and coupling has been constructed with Robin boundary conditions applied \cite{Dappiaggi:2018xvw}, the resulting renormalized expectation values have yet to be computed.

In this paper we study a real, free scalar field with general mass and coupling propagating on global three-dimensional adS, and compute the v.e.v.~and thermal expectation value (t.e.v.) of the VP. 
Our motivations for working on a three-dimensional background are two-fold. 
First, reducing the number of space-time dimensions simplifies the angular part of the scalar field Green's function, thus avoiding the use of conical functions which were required in the analysis in~\cite{Morley:2021}.
Second, our work in this paper can be considered as a prelude to a study of renormalized expectation values on the three-dimensional BTZ black hole \cite{Banados:1992wn,Banados:1992gq}.
Naively one would expect that, far from the black hole, renormalized expectation values would approach those on pure adS. 
We therefore focus particularly on the behaviour of the VP close to the adS boundary. 
 
 We start, in section \ref {sec:DirichletNeumann}, by constructing the vacuum and thermal Feynman's Green's functions for both Dirichlet and Neumann boundary conditions. 
 We then use these Green's functions to compute the v.e.v.~and t.e.v.~of the VP  using Hadamard renormalization.  
 In section \ref{sec:VP_Robin}, following~\cite{Morley:2021}, we employ Euclidean methods to determine the vacuum and thermal Euclidean Green's function which we then use to determine the v.e.v.s and t.e.v.s of the VP with Robin boundary conditions applied to {\em {all}} field modes. 
 The behaviour of these expectation values close to the space-time boundary is examined in more detail in section \ref{sec:boundary}. 
 Our conclusions are presented in section~\ref{sec:conc}.
 
\section{Vacuum polarisation with Dirichlet and Neumann boundary conditions}
\label{sec:DirichletNeumann}

We consider three-dimensional adS in global coordinates, with metric
\begin{equation}
	ds^{2}=L^{2} \sec ^{2} \rho \, \left[ -dt^{2}+d\rho ^{2} + \sin ^{2} \rho \, d\theta ^{2} \right] 
	\label{eq:metric}
\end{equation}
where $0\le \rho <\pi /2$ and $0\le \theta < 2\pi $. The cosmological constant $\Lambda <0$ is related to the inverse radius of curvature $L$ via $\Lambda = -1/L^{2}$. 
For adS, the time coordinate $t$ is periodic with period $2\pi $, and one can take $-\pi \le t \le \pi $ with $t=-\pi $ and $t=\pi $ identified. 
As in \cite{Morley:2021}, here we work on the covering space of adS, denoted CadS, for which the time coordinate is unwrapped, $-\infty  < t < \infty $. 
Our focus in this paper is a real scalar field $\Phi $, of mass $m$, satisfying the Klein-Gordon equation
\begin{equation}
	\left[ g^{\mu \nu }\nabla _{\mu } \nabla _{\nu } - m^{2} - \xi R \right] \Phi =0,
	\label{eq:KG}
\end{equation}
where $R=-6/L^{2}$ is the Ricci scalar curvature.  The coupling constant $\xi $ takes the value $1/8$  for conformal coupling in three dimensions.
To simplify the notation, we define new quantities $\nu $ and $\mu$ by 
\begin{equation}
	\nu = {\sqrt {1+  \mu^2\,L^{2}}}, 
	\label{eq:nu}
\end{equation}
where
\begin{equation}
    \mu^2=m^2 + \xi R.
   \label{eq:musquared}
\end{equation}
In this paper we consider $\nu \in [0,1]$, where $\nu=1/2$ corresponds to the massless, conformally coupled field and $\nu=1$ is the massless, minimally coupled field. In order to satisfy the Breitenlohner-Freedman bound~\cite{Breitenlohner:1982jf}, we require $\nu \ge 0$ for the scalar  field to be classically stable. In~\cite{Ishibashi:2004wx} the possible boundary conditions that can be imposed at the space-time boundary were studied for all values of $\nu$. It was shown that for $\nu \in (0,1)$ many boundary conditions are possible. For $\nu=0$, there is a one-parameter family of possible boundary conditions, whilst for $\nu \ge 1$ only Dirichlet boundary conditions are permitted.

\subsection{Vacuum Green's function and expectation values}
\label{subsec:vacuum_Greens}
Expectation values of the product of field operators can be determined using  Green's functions~\cite{Birrell:1982ix}. We consider the Feynman Green's function, $G_F(x,x')$,  defined as 
\begin{equation}
   G_F(x,x')= i \langle \mathcal{T} (\hat{\Phi}(x)\,\hat{\Phi}(x'))\rangle,
\end{equation}
where $\mathcal{T}$ represents the time ordered product of the field.
The Feynman Green's function satisfies the inhomogenous scalar field equation~\cite{Decanini:2005eg}
\begin{equation}
  \left[ g^{\alpha \beta }\nabla _{\alpha } \nabla _{\beta } - \mu^2 \right] G_F(x,x') = -\frac{1}{\sqrt {|g|}} \delta ^3 (x-x'),
  \label{eq:GF}
\end{equation}
where $g$ is the determinant of the metric~\eqref{eq:metric} and $\delta ^3 (x-x')$ is the three-dimensional Dirac delta function. Owing to the maximum  symmetry of the Green's function when Dirichlet or Neumann  boundary condtions are imposed, in these cases $G_F(x,x')$ depends only on $s(x,x')$, the proper distance between the points $x$ and $x'$ which are connected by a unique geodesic. 

Following the analysis in \cite{Allen}, the general maximally symmetric solution of (\ref{eq:GF}) is
\begin{equation}
G_F (x,x')=C F \left(1+ \nu,1-\nu,\frac{3}{2};\,z\right) + D z^{-\frac{1}{2}} \,F\left(\,\frac{1}{2}+\nu,\frac{1}{2}-\nu,\frac{1}{2};\,z\right)
\label{eq:vacgreen}
\end{equation}
where $F(a,b,c;z)={}_{2}F_{1}(a,b,c;z)$ is a hypergeometric function (the subscripts $2,1$ are omitted for brevity),  $C$ and $D$ are arbitrary constants and 
\begin{equation}
    z= - \sinh^2 \left(\frac{s}{2L} \right).
    \label{eq:variable2}
\end{equation}
This solution is valid for $\nu \in [0,1]$ but for now we focus on $\nu \in (0,1)$ and return to the special cases of $\nu=0,1$ later.  
The constant $D$ is fixed by matching the short-distance behaviour of (\ref{eq:vacgreen}) to that of 
$\widetilde{G}_H$, the divergent part of  the Hadamard parametrix, which is~\cite{Kent:2014nya} 
\begin{equation}
\widetilde{G}_H(x,x') = \frac{i}{4 \pi s}.
\label{eq:divhad}
\end{equation}
This gives
\begin{equation} 
 D =  -\frac{1}{8\pi L}.
\label{eq:Dconstant}
\end{equation}
The constant $C$ is determined by the boundary conditions. 

To apply Dirichlet boundary conditions, we require that $ G_F(s) \to 0$ as rapidly as possible when $ s \to \infty$, yielding
\begin{equation} 
 C=C^{D} = - \frac{i \nu}{4 \pi L}.
\label{eq:Cconstant}
\end{equation}
If the field is massless and conformally coupled ($\nu = 1/2$) we may define Neumann boundary conditions by making a conformal transformation onto the Einstein static universe (ESU) \cite{Avis:1977yn}:
\begin{equation}
  \tilde{g}_{\mu \nu} = \Omega^2 g_{\mu \nu}
\end{equation}
where $ \tilde{g}_{\mu \nu}$ is the metric in ESU, $g_{\mu \nu} $ is the metric (\ref{eq:metric}) in adS and 
\begin{equation} 
  \Omega^2= \cos ^{2}\rho 
  \label{eq:transform}
\end{equation}
is the conformal factor.
If we then impose the requirement that the resulting Green's function on ESU has a derivative which vanishes on the boundary, we find, in this case, that $C=-iD$.
If the scalar field is not conformally invariant, the above method of mapping to ESU no longer applies, and there is a choice to made in how `Neumann' boundary conditions are defined~\cite{Dappiaggi:2018xvw}. For $0 < \nu < 1$, `generalised' Neumann boundary conditions, which correspond to the vanishing of the derivative of the product of the field with trigonometric functions of the radial coordinate can be defined \cite{Ishibashi:2004wx}.  
In keeping with~\cite{Ishibashi:2004wx,Dappiaggi:2018xvw}, we now choose the relationship $C=-iD$ between $C$ and $D$ for the Neumann boundary condition for general $\nu$, yielding 
\begin{equation}
  C=C^{N} =  \frac{i \nu}{4 \pi L}.
  \label{eq:NCconstant}
\end{equation}

The vacuum Green's functions $G_{0}^{D/N}(s)$ with Dirichlet (D) and Neumann (N) boundary conditions applied are then
\begin{align}
   G_0^{D} (s) = & - \frac{i \nu }{4 \pi L}  F\left (1+ \nu,1-\nu,\frac{3}{2};-\sinh^2\left( \frac{s}{2L}\right)\right)
 + \frac{i}{8 \pi L\, \sinh\left(\frac{s}{2L}\right)} F \left(\frac{1}{2}+\nu,\frac{1}{2}-\nu,\frac{1}{2};-\sinh^2\left( \frac{s}{2L}\right)\right),
\label{eq:GFoD} 
\\ 
G_0^{N} (s) = &  \frac{i \nu }{4 \pi L}  F\left (1+ \nu,1-\nu,\frac{3}{2};-\sinh^2\left( \frac{s}{2L}\right)\right) 
+ \frac{i}{8 \pi L\, \sinh\left(\frac{s}{2L}\right)} F \left(\frac{1}{2}+\nu,\frac{1}{2}-\nu,\frac{1}{2};-\sinh^2\left( \frac{s}{2L}\right)\right),
\label{eq:GFoN}
\end{align}
which differ only in the sign of the first term. 
From these, we can determine the renormalised v.e.v.~of the VP,  denoted by  $\langle\hat{\Phi}^2\rangle_{0}$, which is given by
\begin{equation}
 \langle\hat{\Phi}^2\rangle_{0}\,=\lim_{s \to 0}  [-iG_0^{D/N} - (-i \widetilde{G}_H)],
\label{eq:expvalue} 
\end{equation}
where $\widetilde{G}_H$ is the divergent part of  the Hadamard parametrix (\ref{eq:divhad}).
For Dirichlet boundary conditions, the renormalized VP is then \cite{Kent:2014nya}
\begin{equation}
 \langle \hat{\Phi}^2\rangle^D_{0}=- \frac{\nu}{4 \pi L},
 \label{eq:DvacExpvalue}
\end{equation}
while for Neumann boundary conditions we have
\begin{equation}
 \langle\hat{\Phi}^2\rangle^N_0\,=  \frac{\nu}{4 \pi L}.
 \label{eq:VPN}
\end{equation}
This has the same magnitude as the Dirichlet case but opposite  sign. It can be seen that in the case $\nu=0$ both boundary conditions give the same result (see section \ref{subsec:nu0,1} where we consider this case in more detail). With both Dirichlet and Neumann boundary conditions, we note that the v.e.v.s depend only on $\nu$ and are independent of the radial coordinate, as expected from maximal symmetry.

\subsection{Thermal Green's functions and expectation values}
\label{subsec:thermal_Greens}
We now determine the thermal Green's function, $G_\beta (t,{\bm {x}};t',{\bm {x}}')$, for inverse temperature $\beta $, where ${\bm {x}}=(\rho , \theta )$.
The thermal Green's function
can be expressed as an infinite sum involving the vacuum Green's function $G_{0}(x,x')$, as follows \cite{Birrell:1982ix}:
\begin{equation}
G_\beta(t,{\bm{x}};t',{\bm{x}}')=\sum_{j=-\infty}^{\infty} G_{0} (t + i j\beta, {\bm{x}}; t',{\bm{x}}').
\label{thermal_vac}
\end{equation}
As the divergent part of the Hadamard parametrix~\eqref{eq:divhad} is independent of the state and the renormalized v.e.v.~has already been calculated, we do not need to repeat the renormalisation process for the t.e.v.. Instead we can look at the difference between the t.e.v.~and the v.e.v.~using the result
\begin{equation}
\langle{\hat {\Phi}}^2\rangle_\beta- \langle{\hat {\Phi}}^2\rangle_0 = -i\sum_{j=-\infty,j\ne 0}^{\infty} G_\beta (t + i j \beta, {\bm {x}},t ,{\bm {x}}).
\label{eq:thermaleq}
\end{equation}
We obtain the thermal Green's functions for Dirichlet and Neumann boundary conditions from  the vacuum Green's functions~(\ref{eq:GFoD}, \ref{eq:GFoN}) using~\eqref{thermal_vac}. 
The Green's functions appearing in the sum in (\ref{thermal_vac}) depend on the proper distance $s(x,x')$ between the space-time points $(t,{\bm {x}})$ and $(t'+ij\beta ,{\bm {x}}')$.
The proper distance $s(x,x')$ is given by~\cite{Avis:1977yn}
\begin{equation}
\cosh \left( \frac{s}{L}\right) = \frac{\cos \Delta t}{\cos \rho \cos \rho'}- \cos \Delta \theta \tan \rho \tan \rho',
\label{eq:cosh_s}
\end{equation}
for two space-time points $x (t,\rho, \theta)$ and $x' ( t', \rho', \theta')$ in adS with $ \Delta t = t'-t$ and  $\Delta \theta = \theta'-\theta$.
In the limit $t'\rightarrow t$, ${\bm {x}}'\rightarrow {\bm {x}}$, 
we find that the thermal Green's functions in (\ref{eq:thermaleq}) are even functions of $j$, and hence arrive at the result
\begin{multline}
\langle{\hat {\Phi}}^2\rangle^D_\beta- \langle{\hat {\Phi}}^2\rangle^D_0\, = \frac{1}{2 \pi L} \sum_{j=1}^{\infty}  \left\{ -\nu F\left (1+ \nu,1-\nu,\frac{3}{2};\frac{1-\cosh(j\beta)}{2\cos^2\rho}\right) \right. \\
 \left.+ \frac{\sqrt{2}\cos\rho}{2 \sqrt{\cosh(j\beta)-1}} F \left(\frac{1}{2}+\nu,\frac{1}{2}-\nu,\frac{1}{2};\frac{1-\cosh(j\beta)}{2\cos^2\rho}\right) \right \},
\label{eq:thermalD}
\end{multline}
for the Dirichlet boundary condition, while for the Neumann boundary condition we have
\begin{multline}
\langle{\hat {\Phi}}^2\rangle^N_\beta- \langle{\hat {\Phi}}^2\rangle^N_0\, = \frac{1}{2 \pi L} \sum_{j=1}^{\infty}  \left\{ \nu F\left (1+ \nu,1-\nu,\frac{3}{2};\frac{1-\cosh(j\beta)}{2\cos^2\rho}\right) \right. \\
 \left.+ \frac{\sqrt{2}\cos\rho}{2 \sqrt{\cosh(j\beta)-1}} F \left(\frac{1}{2}+\nu,\frac{1}{2}-\nu,\frac{1}{2};\frac{1-\cosh(j\beta)}{2\cos^2\rho}\right) \right \}.
\label{eq:thermalN}
\end{multline}

\begin{figure}[htbp]
     \begin{center}
        \includegraphics[width=0.47\textwidth]{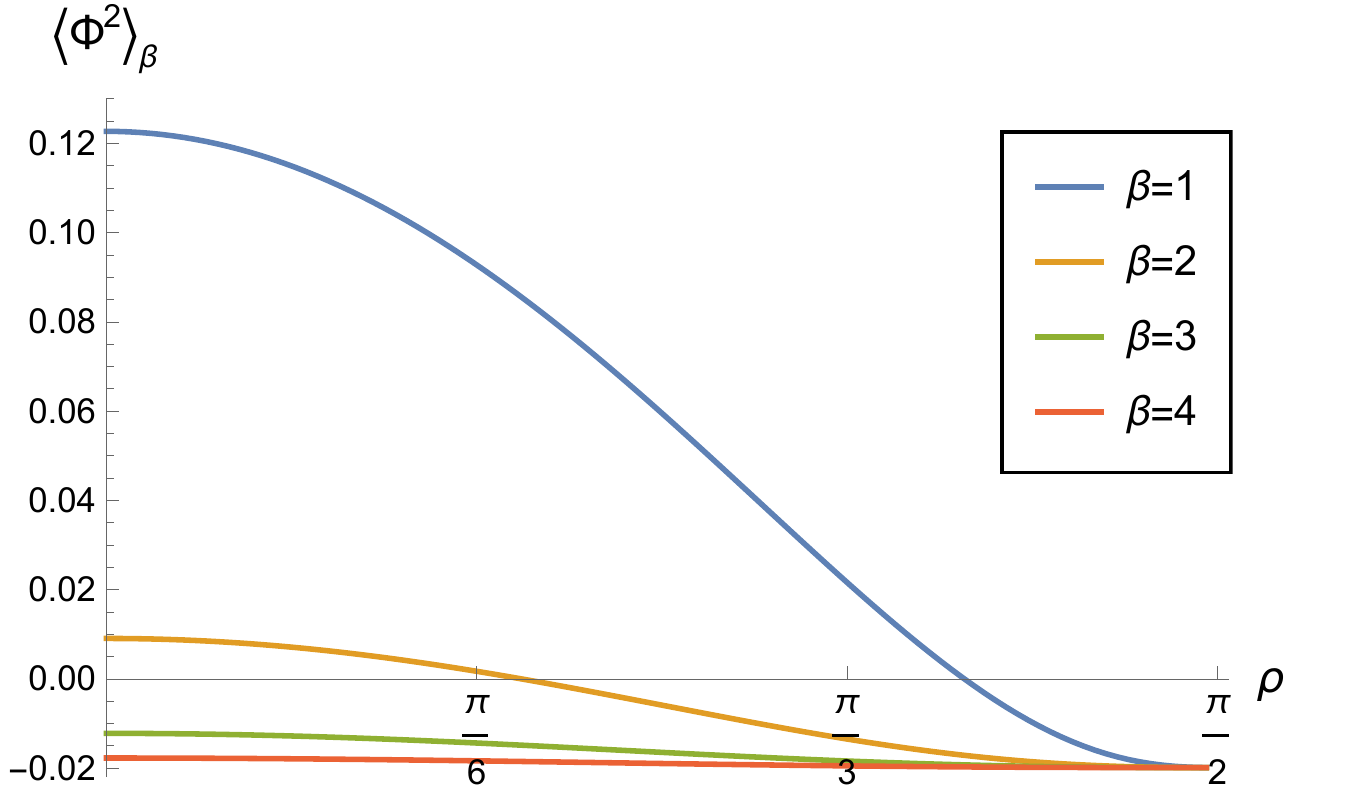} \quad
       \includegraphics[width=0.42\textwidth]{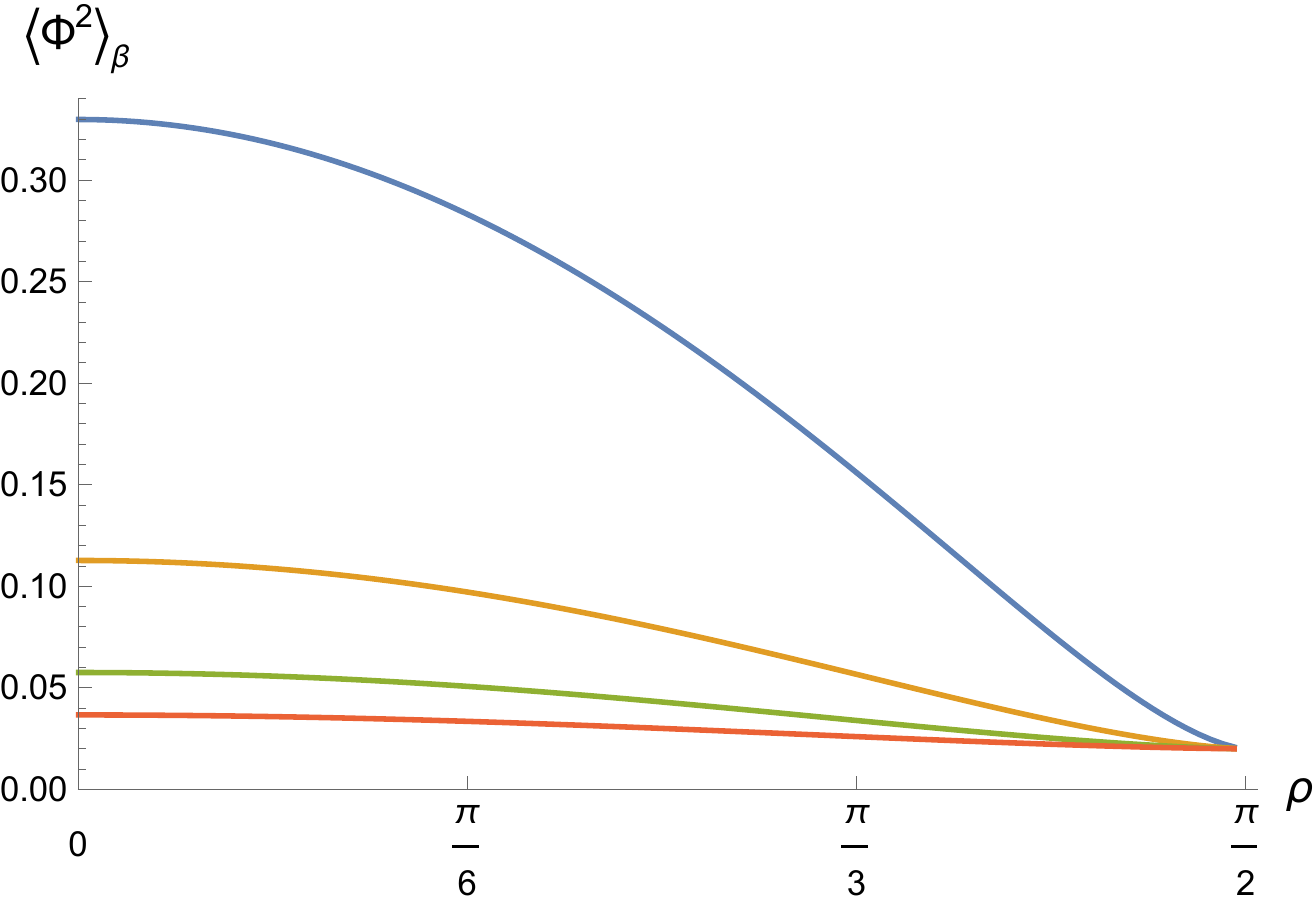}
        \caption{Renormalized t.e.v.s of the VP $\langle \hat{ \Phi}^2 \rangle _\beta$ with Dirichlet  (left) and Neumann (right) boundary conditions for the scalar field with  $\nu=1/4$ and four different values of the inverse temperature $\beta$.}
        \label{fig:thermal}
      \end{center}
   \end{figure}

\begin{figure}[htbp]
     \begin{center}
        \includegraphics[width=0.47\textwidth]{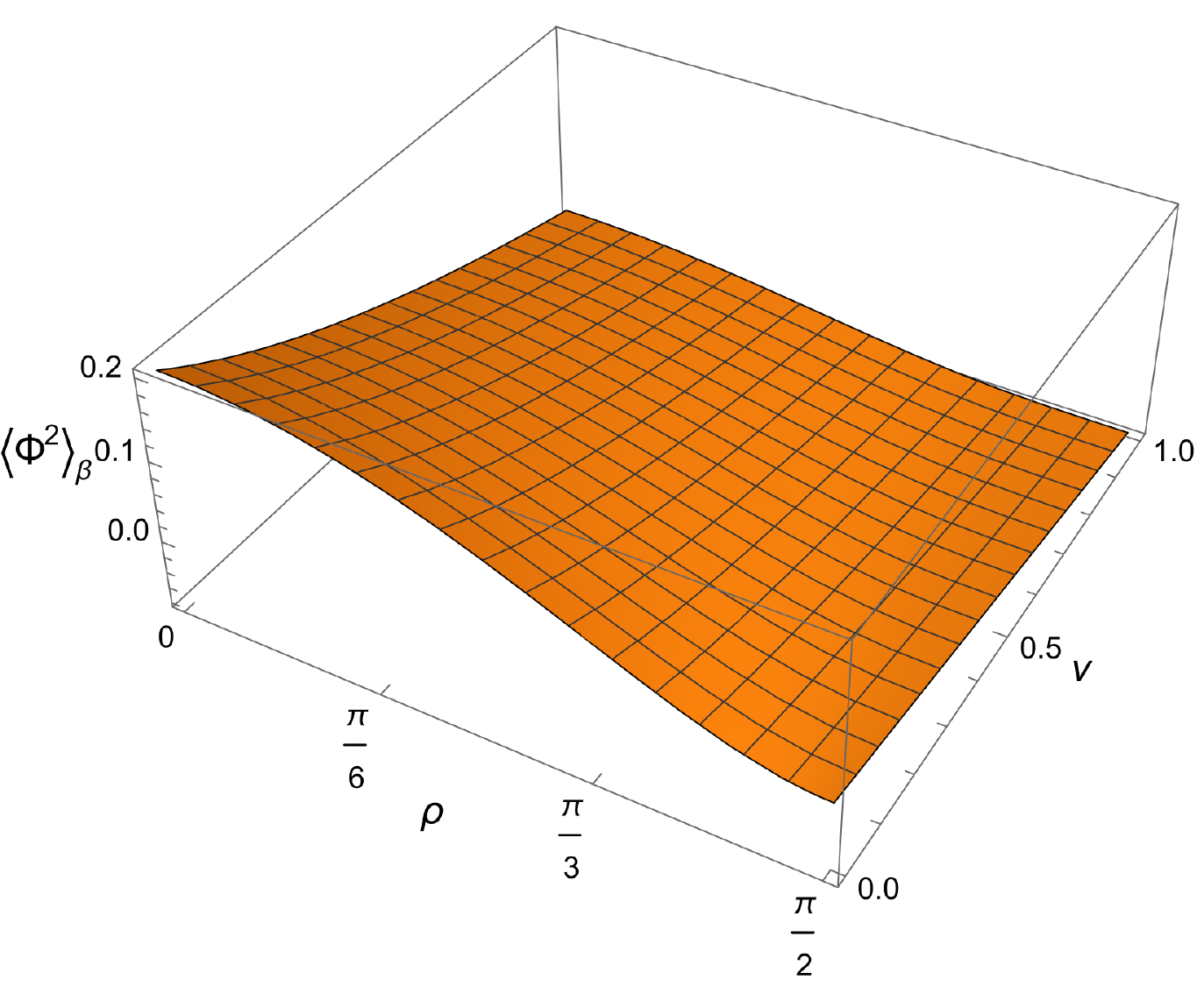}\quad
        \includegraphics[width=0.47\textwidth]{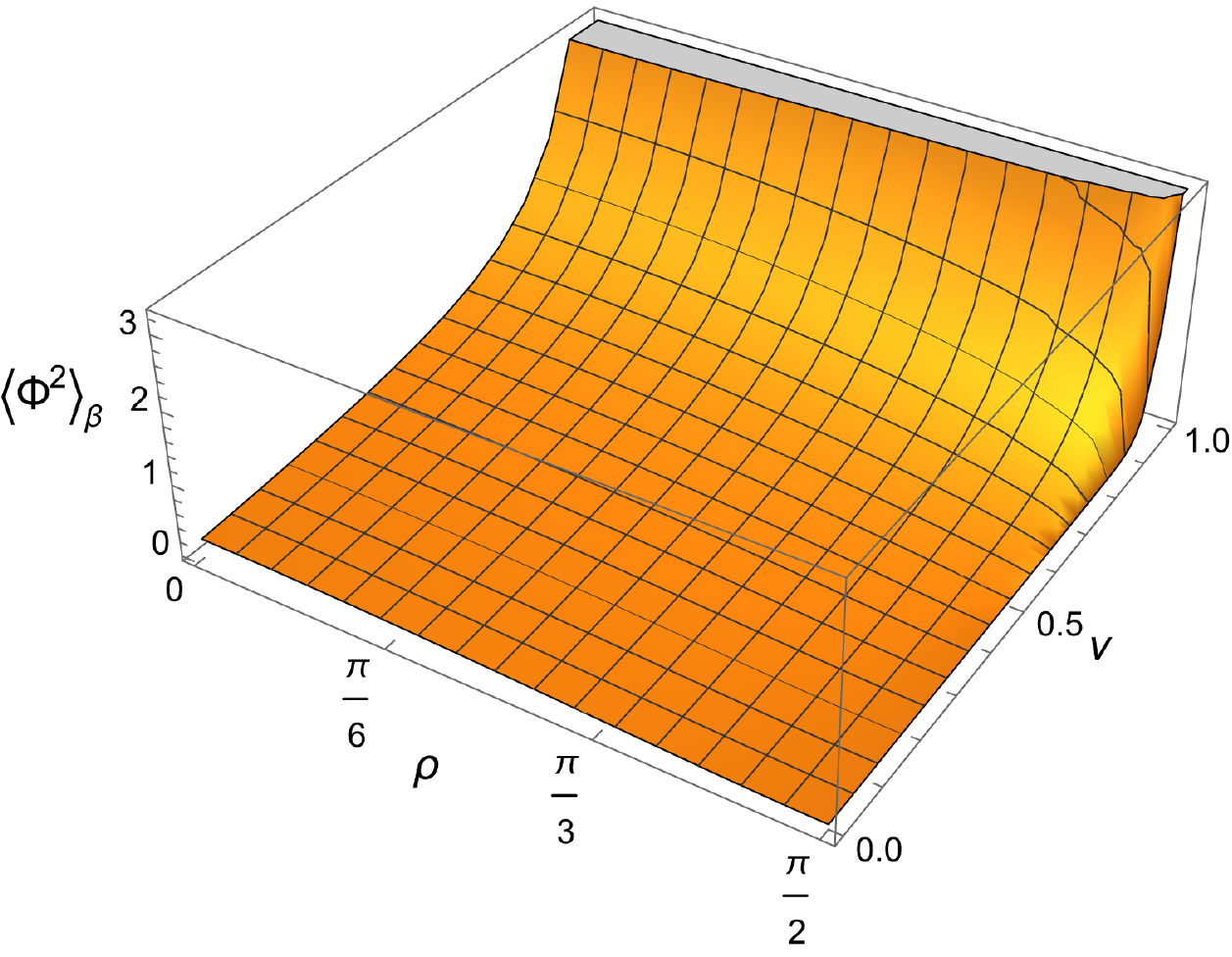}
        \caption{T.e.v.~of the VP $\langle \hat{\Phi}^2 \rangle _\beta$, as a function of $\rho$ and $ \nu$,  with Dirichlet  (left) and Neumann  (right) boundary conditions. In both  cases $\beta=1$.}
        \label{fig:surface}
      \end{center}
   \end{figure}

\begin{figure}[htbp]
     \begin{center}
        \includegraphics[width=0.47\textwidth]{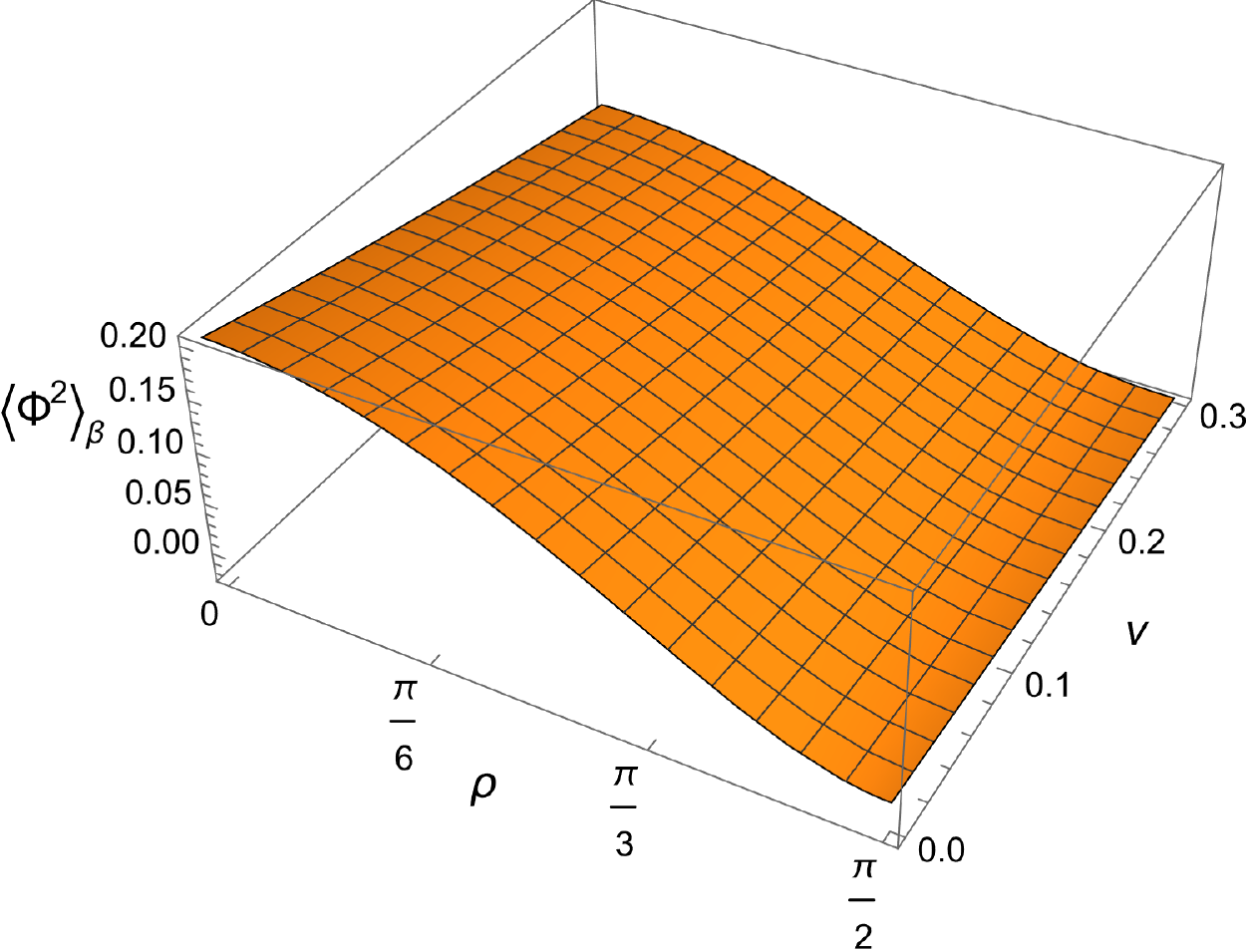}\quad
        \includegraphics[width=0.47\textwidth]{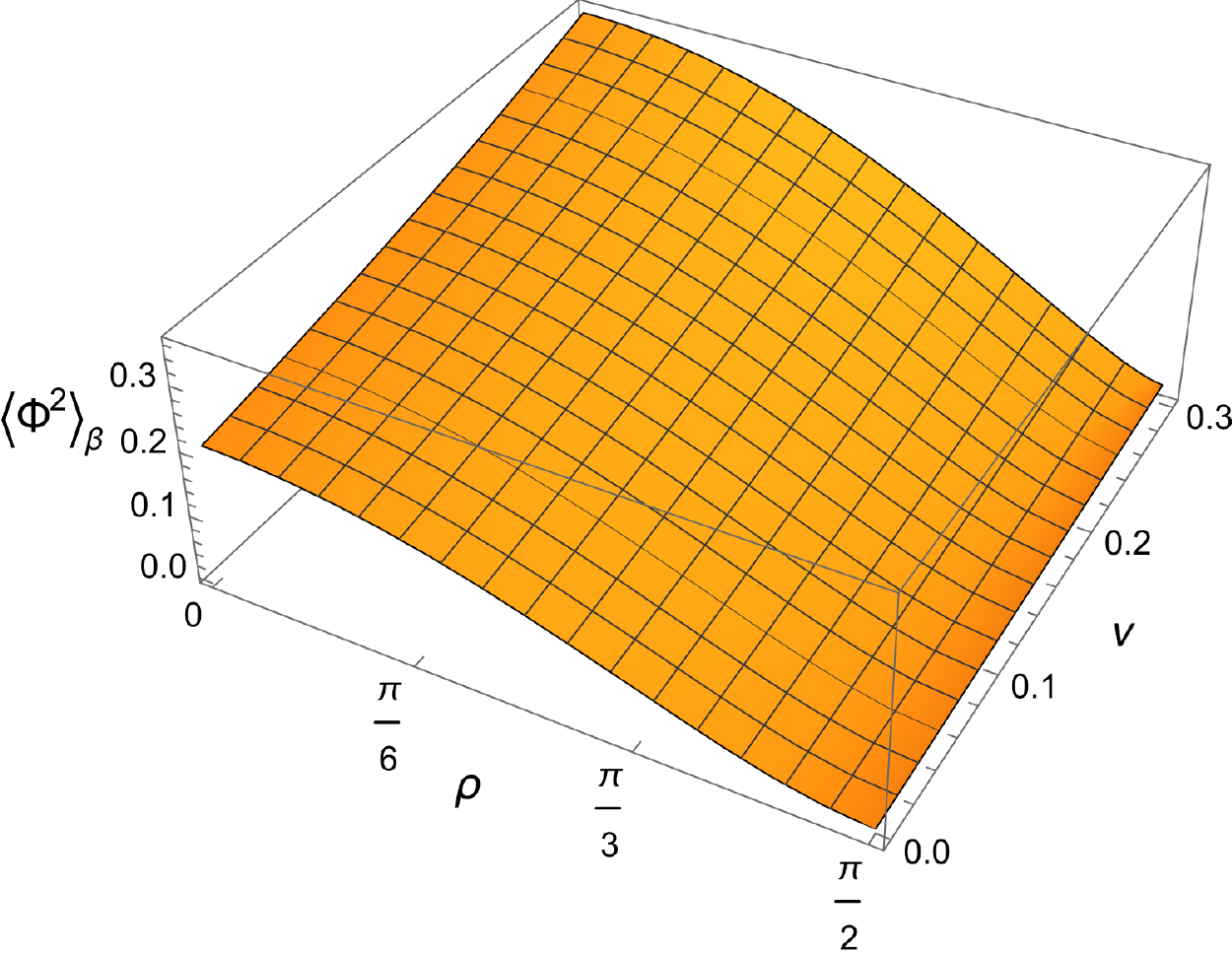}
        \caption{Close up view of figure~\ref{fig:surface} showing $\langle \hat{ \Phi}^2 \rangle _\beta$, for Dirichlet  (left) and Neumann  (right) boundary conditions, for $\nu \in (0, 0.3)$ and $\beta=1$.}
        \label{fig:close}
      \end{center}
   \end{figure}

The thermal expectation values, $ \langle \hat{\Phi}^2\rangle_\beta$, with both Dirichlet and Neumann boundary conditions (\ref{eq:thermalD}, \ref{eq:thermalN}) are calculated numerically using {\footnotesize MATHEMATICA}. We find that the sums in  (\ref{eq:thermalD}, \ref{eq:thermalN}) converge rapidly. The results are shown in figures~\ref{fig:thermal}--\ref{fig:close} for different values of the inverse temperature $\beta$.
   
Figure~\ref{fig:thermal} shows $ \langle\hat{\Phi}^2\rangle_\beta$  for Dirichlet (left) and Neumann (right) boundary conditions, for four different values of the inverse temperature $\beta$ and a representive value of $\nu=1/4$. 
Figure~\ref{fig:surface} shows the  three-dimensional surface plots of  $ \langle \hat{\Phi}^2\rangle_\beta$ as a function of $\rho$ and $\nu$, again for Dirichlet (left) and Neumann (right) boundary conditions. 
The plots in figure~\ref{fig:thermal} are similar to those in 
\cite{Ambrus:2018olh}, showing that the thermal energy of the field concentrates in a region close to the origin, thereby breaking the translational symmetry of the background adS space-time.
For all values of the inverse temperature and $\nu \in (0,1)$, the t.e.v.~of the VP attains its maximum value at the origin, and is monotonically decreasing towards the boundary.
We also see that for increasing $\beta$ (decreasing temperature), the t.e.v.s approach the v.e.v.s, as observed in \cite{Morley:2021} for the four-dimensional, massless, conformally coupled scalar field. 
Approaching the  space-time boundary as $\rho\rightarrow\pi/2$, we recover the zero temperature v.e.v.~for all $\beta$ and $\nu$. 
For the Neumann boundary condition, the convergence to the vacuum result, as $\rho \to \pi/2$, occurs less rapidly than in  the Dirichlet case. 

Although the t.e.v.s for the cases of $ \nu= 0,1$ have not specifically been determined in the above analysis (see section \ref{subsec:nu0,1} for analysis of these values of $\nu $), figure~\ref{fig:surface} suggests that $ \langle \hat{\Phi}^2\rangle_\beta$ reaches the same value for both Dirichlet and Neumann boundary conditions as $\nu \to 0$ for all $\rho$. 
This can be seen more clearly in figure~\ref{fig:close}, which shows an expanded view of figure~\ref{fig:surface} for $\nu \in (0, 0.3)$.  For the Dirichlet boundary condition, at fixed $\rho $ the t.e.v.~ $ \langle \hat{\Phi}^2\rangle_\beta$ has its  maximum value at $ \nu=0$ and decreases for increasing $ \nu$. For the Neumann boundary condition, however, at fixed $\rho $ the t.e.v.~$ \langle \hat{\Phi}^2\rangle_\beta$ has its minimum value at $\nu=0$, and increases steadily for increasing $\nu$.
As $\nu\rightarrow 1$, it appears to be the case that $ \langle \hat{\Phi}^2\rangle_\beta$ diverges, in accordance with the result in~\cite{Ishibashi:2004wx} that there is no generalized Neumann boundary condition for $\nu=1$. 

\subsection{Vacuum and thermal expectation values for $\nu=0,1$}
\label{subsec:nu0,1}
The analysis in sections~\ref{subsec:vacuum_Greens} and \ref{subsec:thermal_Greens} was valid for $\nu \in (0,1)$. We now consider the particular cases  $\nu=0,1$. 

First, for $\nu=0$, the vacuum Green's function \eqref{eq:vacgreen} simplifies to  
\begin{equation}
  G_F (x,x')=\frac{C \sin^{-1}\sqrt{z}}{\sqrt{1-z}\,\sqrt{z}} - \frac{1}{8\pi L\sqrt{z}\sqrt{1-z}},
\label{eq:greenv0}
\end{equation} 
where we have used (\ref{eq:Dconstant}). 
To obtain the value of the constant $C$, we consider the form of the Green's function \eqref{eq:greenv0} at the space-time boundary. Using~\eqref{eq:variable2} and the series  expansion for $\sin^{-1} \sqrt{z}$ as $ z \to -\infty$ ($s \to \infty$),  
we find that $G_F(s) \to 0$  as $s \to \infty$ for all values of $C$. 
However, applying Dirichlet boundary conditions and demanding rapid convergence to zero requires $C=0$, as in (\ref{eq:Cconstant}). 
Applying (\ref{eq:NCconstant}) for $\nu =0$, we would find that $C=0$ for Neumann boundary conditions also.
For this value of $\nu $ our earlier definition of Neumann boundary conditions therefore corresponds to Dirichlet boundary conditions, and it makes no sense to consider Robin (mixed) boundary conditions.
However, in this case there remains a one-parameter family of boundary conditions leading to consistent classical dynamics~\cite{Ishibashi:2004wx}.

We can straightforwardly construct a one-parameter family of maximally symmetric vacuum Green's functions simply by choosing a non-zero value of the constant $C$. In order to obtain a real expectation value for the v.e.v., we define $C =i\widetilde{C}$ for~$\widetilde{C} \in \mathbb{R}$. 
Using the Maclaurin series expansions of the functions in (\ref{eq:greenv0}), the renormalized v.e.v.~of the VP is then
 \begin{equation}
    \langle\hat{\Phi}^2\rangle_0= \widetilde{C}. 
    \label{eq:v0vacD}
 \end{equation}
Varying the constant ${\widetilde {C}}$ therefore results simply in a constant shift of the v.e.v.~of the VP.
We note that we could similarly define a one-parameter family of maximally symmetric Green's functions for other values of $\nu $, simply by allowing the constant $C$ in (\ref{eq:vacgreen}) to vary.
However, the physical interpretation of the states given by such maximally symmetric Green's functions is unclear; in particular they do not correspond to the application of Robin boundary conditions \cite{Pitelli:2019svx}.
We therefore do not consider such states further, except when $\nu =0$ and the usual Robin boundary conditions are not applicable.

For $\nu=1$, 
we can write the vacuum  Feynman Green's function \eqref{eq:GFoD}, with Dirichlet boundary condition, as 
\begin{equation}
   G_0^{D} (s) = -\frac{i}{4\pi L} + \frac{i\cosh (\frac{s}{L})}{8 \pi L \sinh^2(\frac{s}{2L})\cosh^2(\frac{s}{2L})}.
   \label{eq:DGFv1}
\end{equation}
 Using (\ref{eq:expvalue}, \ref{eq:DGFv1}) gives the v.e.v.~of the VP, with Dirichlet  boundary conditions, to be $ -1/4 \pi L$, consistent with the value found for $\nu \in (0,1)$ in \eqref{eq:DvacExpvalue}.

\begin{figure}[htbp]
     \begin{center}
        \includegraphics[width=0.9\textwidth]{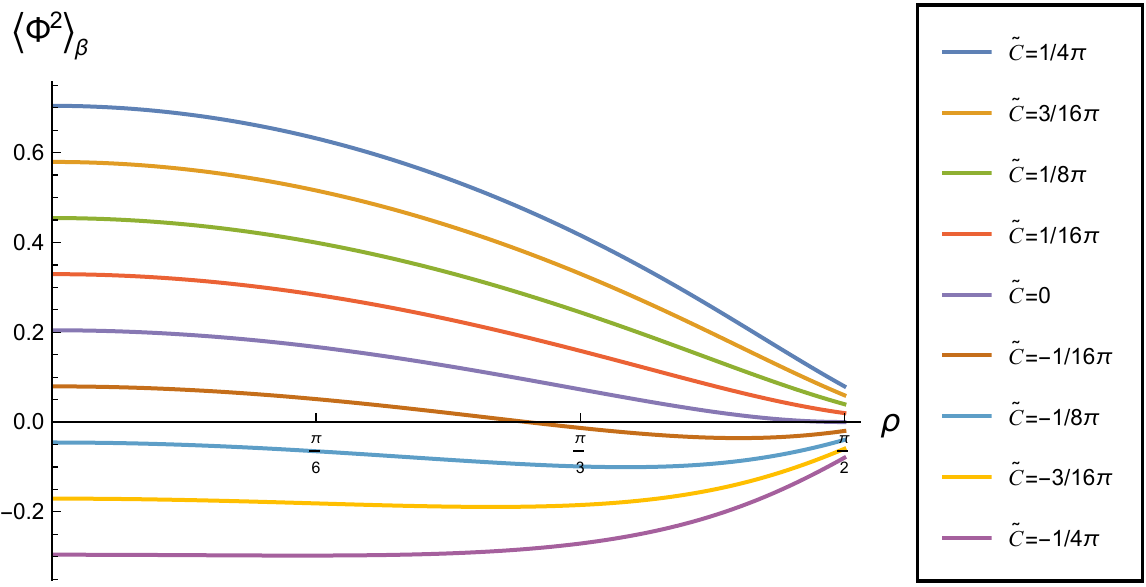} 
               \caption{T.e.v.s of the VP as functions of the radial coordinate $\rho$, for $\nu=0$ and different values of $ \tilde{C}$ with $\beta =1$. }
        \label{fig:thermal_v0}
      \end{center}
\end{figure}

We can calculate the t.e.v.~for the VP, for $\nu=0,1$, along the same lines as we did for $\nu \in (0,1)$ in section~\ref{subsec:thermal_Greens}. 
For $\nu=0$, 
we express the difference between the  t.e.v.~and the v.e.v.~of the VP as 
\begin{equation}
\langle \hat{\Phi}^2\rangle_\beta -\langle \hat{\Phi}^2\rangle_0=\sum_{j=1}^{j= \infty} \Bigg \{ \frac{ 2 \sqrt{2} \,\widetilde{C} \cos \rho}{\sqrt{\cosh j \beta + \cos 2\rho}}  
+\frac{ \cos^2 \rho}{2 \pi L \sqrt{(\cosh j\beta -1)(\cosh j \beta +\cos 2\rho)}} \Bigg\},
\end{equation}
where $\langle \hat{\Phi}^2\rangle_0$ is given in~\eqref{eq:v0vacD}. Figure~\ref{fig:thermal_v0} shows the  t.e.v.s for various different values of $\widetilde{C}$ and inverse temperature $\beta =1$. As we approach  the space-time boundary ($\rho \to \pi/2)$, the t.e.v.s converge to the v.e.v.s (which depend on $\widetilde{C}$ \eqref{eq:v0vacD}). For $\widetilde{C}=0$, the t.e.v.s match those seen in figure~\ref{fig:close} for $\nu \to 0$. 
One interesting feature of the t.e.v.s in figure \ref{fig:thermal_v0} is that the profiles of the t.e.v.s are no longer always monotonically decreasing functions of the radial coordinate $\rho $.
When ${\widetilde {C}}>0$, it remains the case that the t.e.v.s have their maximum values at the origin and decrease towards the boundary. However, for ${\widetilde {C}}<0$, the t.e.v.s are monotonically increasing close to the boundary. 
When ${\widetilde {C}}$ is small and negative, there is still a maximum at the origin, but for sufficiently large $\lvert {\widetilde {C}} \rvert $ the t.e.v.s have a minimum at the origin and are monotonically increasing as $\rho $ increases.

Using the same method for $\nu=1$, the t.e.v. of the VP, with Dirichlet boundary conditions applied, is found to be 
\begin{equation}
\langle \hat{\Phi}^2\rangle ^D_\beta- \langle \hat{\Phi}^2\rangle ^D_0 = \frac{1}{2 \pi L} \sum_{j=1}^{j=\infty} \left \{ \frac{(\cosh j\beta - \sin^2 \rho)}{\sqrt{\cosh j\beta-1}\sqrt{\cosh j\beta + \cos 2\rho}}-1 \right \},
\label{eq:Dthermal}
\end{equation}
from which it can be readily seen that at $\rho=\pi/2$, we obtain $\langle \hat{\Phi}^2\rangle ^D_\beta = - 1/ 4\pi L$  consistent with  figure~\ref{fig:surface}. 
If we try to find the t.e.v.~of the VP with Neumann boundary conditions (by using \eqref{eq:thermalN} for $v=1$), we obtain 
\begin{equation}
\langle \hat{\Phi}^2\rangle ^N_\beta- \langle \hat{\Phi}^2\rangle ^N_0 = \frac{1}{2 \pi L} \sum_{j=1}^{j=\infty} \left \{ \frac{(\cosh j\beta - \sin^2 \rho)}{\sqrt{\cosh j\beta-1}\sqrt{\cosh j\beta + \cos 2\rho}} +1 \right \}.
\label{eq:cosh_div}
\end{equation}
The first term in brackets in  \eqref{eq:cosh_div} will tend to unity for large $j$ showing that the t.e.v.~in this case is divergent, again consistent with figure~\ref{fig:surface}. 

\section{Vacuum polarisation with Robin boundary conditions}
\label{sec:VP_Robin}

Having computed the renormalized t.e.v.s and v.e.v.s of the VP for a scalar field with general mass and coupling, when either Dirichlet or Neumann boundary conditions are applied, we now turn our attention to more general mixed (Robin) boundary conditions. 
As in \cite{Morley:2021}, we employ Euclidean methods, so that the Green's function is unique and to circumvent the requirement to use an $i\epsilon$-prescription in order to obtain the Green's function as a well-defined distribution. 

\subsection{Euclidean Green's functions}
\label{subsec:Euclid_Greens}

The Euclidean  metric is obtained from the adS metric 
\eqref{eq:metric} by performing a Wick rotation, $t \to i \tau$, which gives
\begin{equation}
ds^2 = L^2\sec^2\rho \, \,[d\tau^2 + d\rho^2 +\sin^2\rho \, d\theta^2].
\label{eq:Euclid_metric}
\end{equation}
The Euclidean Green's function, $G_E(x,x')$, for the scalar field satisfies the inhomogeneous PDE  
\begin{align}
\underbrace{\left( \Box_E - \mu^2 \right)}_{\mathcal{L}_x} G_E(x,x' ) &= -\frac{1}{\sqrt{g}}\delta^3 ( x-x')
\label{eq:inhomgreen}
\end{align}
where $\mu^2$ is given in \eqref{eq:musquared}, $g$ is the determinant of the Euclidean adS metric \eqref{eq:Euclid_metric} and $\mathcal{L}_x$ is a differential operator whose form is
\begin{equation}
\mathcal{L}_x = \frac{1}{L^2 \sec^2 \rho} \left(  \frac{\partial ^2}{\partial \tau^2}  + \frac{\partial ^2}{\partial \rho^2} + \frac{1}{\text{sin}^2 \rho}
\frac{\partial ^2}{\partial \theta^2} + \frac{1}{\text{sin} \rho \,\, \text{cos} \rho} \frac{\partial}{\partial \rho} -\mu^2L^2 \sec ^2\rho\right).
\end{equation}
Suitable ansatze for the vacuum Euclidean Green's function, $G_E^0(x,x')$ and thermal Euclidean Green's function $G_E^\beta(x,x')$ are
\begin{align}
G_E^0(x,x') & =\frac{1}{4\pi^2} \int_{-\infty}^{\infty} e^{i\omega \Delta \tau} d\omega \,\sum_{\ell=-\infty}^{\infty} e^{i \ell\Delta \theta} g_{\omega\ell} (\rho,\rho'),
\label{eq:Euclid_vac_Green}
\\
G_E^\beta(x,x') & = \frac{\kappa}{4\pi^2} \sum_{n=-\infty}^{\infty} \,\sum_{\ell=-\infty}^{\infty} e^{i n\kappa \Delta \tau}e^{i \ell\Delta \theta} g_{n\ell} (\rho,\rho'),
\label{eq:Euclid_Thermal_Green}
\end{align}
where 
\begin{equation}
  \kappa= \frac{2\pi}{\beta},
  \label{eq:kappa}
  \end{equation}
  and
$ g_{\omega \ell (\rho,\rho')} $ is the vacuum radial Green's function. The thermal radial Green's function $g_{n\ell }(\rho , \rho')$ is given by setting $\omega = n\kappa $ in $g_{\omega \ell} (\rho , \rho')$. Using (\ref{eq:inhomgreen}),  the equation satisfied by $ g_{\omega \ell}(\rho,\rho') $ takes the form
\begin{multline}
 \cos^2 \rho\,\frac{\partial^2 g_{\omega\ell}}{\partial\rho^2} \,(\rho,\rho') + \cot \rho\,\frac{\partial g_{\omega\ell}}{\partial\rho}(\rho,\rho')\\
 + (-\omega^2 \cos^2 \rho - \ell^2 \cot^2 \rho - \mu^2 L^2) g_{\omega\ell}(\rho, \rho')
=-\frac{\cot \rho \cos^2\rho}{L} \delta(\rho-\rho').
\label{eq:inhomeq2}
\end{multline}

The general  solution  $q_{\omega \ell}$ of the homogeneous version of \eqref{eq:inhomeq2} is 
\begin{equation}
\label{eq:g0}
q_{\omega\ell}(\rho)  = [\cos\rho]^A [\sin\rho]^B  \left \{C_1 F (a,b,c;\cos ^{2}\rho )  + C_2\,\, (\cos \rho )^{2(1-c)} F ( a -c +1, b -c + 1, 2-c;\cos ^{2}\rho ) \right \},
\end{equation}
where $C_{1}$ and $C_{2}$ are arbitrary constants, and $F(a,b,c;\cos ^{2}\rho) $
is a hypergeometric function with parameters
\begin{equation}
a = \frac{1}{2}(A+B) - \,\frac{1}{2} i \omega, \qquad
b = \frac{1}{2}(A+B) + \,\frac{1}{2} i \omega, \qquad 
c = 1 + \nu,
\label{eq:alphabetaEuc}
\end{equation}
and
\begin{equation}
A = 1 + \nu, \quad B =  \lvert \ell \rvert \quad \text{for} \quad \ell \in \mathbb{Z}. 
\label{eq:ABEuc}
\end{equation}
Since we are considering $\nu \in (0,1)$, the solution (\ref{eq:g0}) is regular at $\rho = \pi/2$ for all values of $C_{1}$, $C_{2}$.
This is a manifestation of the requirement to impose boundary conditions on the scalar  field at $\rho =\pi /2$. 
We adopt the approach in \cite{Dappiaggi:2018xvw} and define the arbitrary constants $C_{1}$ and $C_{2}$ in terms of a parameter $\zeta \in [0,\pi)$ (hereafter called the `Robin parameter') as follows:
\begin{equation}
C_1=\cos\zeta \quad\text{and}\quad C_2=\sin\zeta. 
\end{equation}
Dirichlet boundary conditions correspond to $\zeta = 0$ and Neumann boundary conditions to $\zeta = \pi /2$.

In general $q_{\omega \ell }(\rho )$ is divergent at the origin $\rho =0$.
The solution $p_{\omega \ell}$ of the homogeneous version of \eqref{eq:inhomeq2} which is regular at the origin is 
\begin{equation}
p_{\omega\ell}(\rho)  = [\cos\rho]^A [\sin\rho]^B\, F ( a, b, a + b - c +1 ; \sin ^{2}\rho ),
\label{eq:g1}
\end{equation}
where we have set an overall arbitrary constant to unity.
The second, linearly independent, solution of the radial equation has a logarithmic singularity at $\rho=0$ \cite[15.10(i)]{NIST:DLMF} and hence we eliminate this solution.
The hypergeometric functions in (\ref{eq:g0}, \ref{eq:g1}) are related by Kummer's connection formulae~\cite[15.10(ii)]{NIST:DLMF}. 
These enable us to write $p_{\omega\ell}(\rho)$ \eqref{eq:g1} in the alternative form 
\begin{equation}
p_{\omega \ell}(\rho ) =  \,[\cos \rho]^A [\sin \rho]^B  \{ \mathcal{P} F( a, b, c; \cos ^{2}\rho )
 +  \mathcal{Q}\,\, (\cos \rho )^{2(1-c)}
 F ( a -c +1, b -c + 1, 2-c;\cos ^{2}\rho ) \},
\label{genrad}
\end{equation}
where 
\begin{equation}
\mathcal{P}= \frac{\Gamma (1- c) \Gamma( a + b - c +1)}{\Gamma ( a - c + 1) \Gamma ( b - c + 1)},
\qquad
\mathcal{Q}=  \frac{\Gamma(c-1) \Gamma ( a + b -c +1)}{\Gamma(a)\Gamma(b)},
\label{eq:pandq}
\end{equation}

 The vacuum radial Green's function, $g_{\omega \ell}(\rho,\rho')$, can be constructed from $q_{\omega\ell}(\rho)$ and $p_{\omega \ell}(\rho)$ by
\begin{equation}
g_{\omega \ell}(\rho,\rho') = \mathcal{N}_{\omega \ell}\, p_{\omega \ell} (\rho_<)\,q_{\omega\ell} (\rho_>)
\label{eq:greencon}
\end{equation}
where  $\rho_< = \min \{ \rho,\rho'\}$, $\rho_> = \max \{ \rho,\rho' \}$ and $ \mathcal{N}_{\omega\ell}$ is a normalisation constant, given by
\begin{equation}
\mathcal{N}_{\omega \ell }=
\frac{1}{2L(1-c)\left[ {\mathcal {P}} \sin \zeta - {\mathcal {Q}} \cos \zeta  \right]}.
\label{eq:NnlR}
\end{equation}
The thermal radial Green's function $g_{n\ell }(\rho , \rho')$ is given by substituting $\omega = n\kappa $ in (\ref{eq:g0}, \ref{eq:g1}, \ref{eq:greencon}, \ref{eq:NnlR}).

The denominator of (\ref{eq:NnlR}) becomes zero if $\zeta $ satisfies 
 \begin{equation}
\tan\zeta =  \frac {\mathcal {Q}}{\mathcal {P}} =  \frac{ \,\Gamma(\nu)\lvert \Gamma(\frac{\lvert \ell \rvert}{2} +\frac{1}{2} -\frac{\nu}{2}-\frac{i\omega}{2})\rvert ^2}{\Gamma(-\nu)\lvert \Gamma(\frac{\lvert \ell \rvert }{2} +\frac{1}{2} +\frac{\nu}{2}-\frac{i\omega}{2})\rvert ^2},
\label{eq:stability}
\end{equation}
resulting in divergent values of~\eqref{eq:NnlR} and a divergent Euclidean Green's function. 
For $\nu \in (0,1)$, the right-hand-side of \eqref{eq:stability} is negative, corresponding to values of $\zeta $ in the interval $(\pi /2,\pi )$.
As in the corresponding quantity for a four-dimensional, massless, conformally coupled scalar field \cite{Morley:2021}, it can be shown that the minimum of the right-hand-side of (\ref{eq:stability}) occurs at $\omega=0$ for fixed $\ell $ and $\nu $ and that it increases monotonically as $\omega$ increases for fixed $\ell $, $\nu $. 
When $\omega =0$, the right-hand-side of \eqref{eq:stability} is an increasing function of $\lvert \ell  \rvert $ for fixed $\nu $.
Therefore the minimum of the right-hand-side of \eqref{eq:stability}, for fixed $\nu $, occurs when $\omega =0 = \ell $.
Substituting $\omega =0=\ell $ into \eqref{eq:stability} gives a value of $\zeta $ which we denote by $\zeta _{\text {crit}}$.
If $0<\zeta < \zeta _{\text{crit}}$, there are no real values of $\omega $ for which \eqref{eq:stability} is satisfied. If $\zeta _{\text{crit}}<\zeta <\pi $, then \eqref{eq:stability} has solutions for $\ell$, $\omega $ and the scalar field is classically unstable \cite{Ishibashi:2004wx}.

\begin{figure}[htbp]
     \begin{center}
        \includegraphics[width=0.8\textwidth]{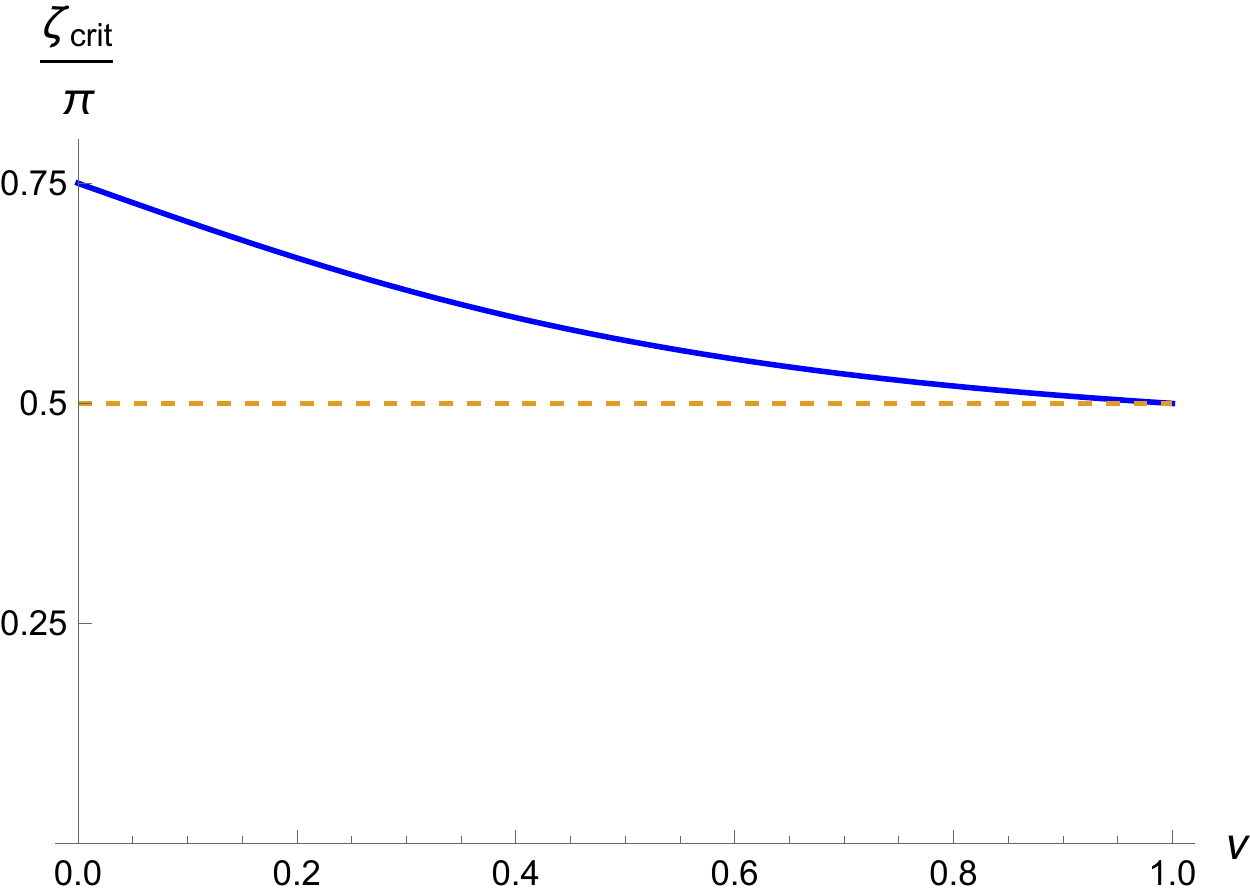} 
               \caption{Critical value $\zeta_{\text{crit}}$ of the Robin parameter as a function of $\nu$, showing that $\zeta_{\text{crit}} \to \pi/2$ as $v \to 1$.}
        \label{fig:zeta_crit}
      \end{center}
\end{figure}

Figure~\ref{fig:zeta_crit} shows $\zeta_{\text{crit}}$ as a function of $\nu$.  In our subsequent computation of v.e.v.s and t.e.v.s, we will only consider values of the Robin parameter $\zeta $ below $\zeta _{\text{crit}}$. 
As $\nu \to 0$, we have $\zeta _{\text{crit}}\to 3\pi /4$. 
In the case of the massless conformally coupled field, for $\nu=1/2$, we find the value $\zeta _{\text{crit}}\approx 0.572 \pi $. As $\nu \to 1$, $\zeta_{\text{crit}} \to \pi/2$, in agreement with the fact that  we cannot consider Neumann boundary conditions for $\nu=1$ (see section~\ref{subsec:nu0,1} and figure~\ref{fig:surface}). 

\subsection{Vacuum expectation values}
\label{subsec:Robin_vac}

Using the results of previous subsection, the vacuum Euclidean Green's function with Robin boundary conditions $g_{\omega\ell}^\zeta (\rho,\rho')$ is 
\begin{align}
g_{\omega\ell}^\zeta &(\rho,\rho') =\mathcal{N}_{\omega\ell}\,\, [\cos\rho]^{1+\nu}\,\, [\cos\rho']^{1+\nu}[\sin\rho]^{\lvert \ell \rvert }\,\,[\sin\rho']^{\lvert \ell \rvert } \nonumber\\
& \times F \Big (\frac{1}{2}(1+\lvert \ell \rvert +\nu - i\omega) ,\frac{1}{2}(1+\lvert \ell \rvert +\nu + i\omega) , 1+\lvert \ell \rvert ; \sin^2\rho _{<}\Big )  \nonumber \\
& \times   \Bigg \{\cos\zeta \,F \Big (\frac{1}{2}(1+\lvert \ell \rvert +\nu - i\omega),\frac{1}{2}(1+\lvert \ell \rvert +\nu + i\omega), 1+\nu; \cos^2\rho_{>} \Big )   \nonumber\\
& +\sin\zeta\,\, [\cos\rho _{>}]^{-2\nu} F \Big (\frac{1}{2}(1+\lvert \ell \rvert -\nu - i\omega),\frac{1}{2}(1+\lvert \ell \rvert -\nu + i\omega), 1-\nu; \cos^2\rho_{>} \Big ) \Bigg \}.
\label{eq:gR}
\end{align}
To calculate the v.e.v.~with Robin boundary conditions, we note that the divergent part of the Hadamard parametrix \eqref{eq:divhad} is state-independent. Thus we simply  consider  the difference between the unrenormalised v.e.v.s with Robin and Neumann boundary conditions, which  removes the singularities common to both. Following this, we  add the v.e.v.~with Neumann boundary conditions \eqref{eq:VPN}. 
We can therefore write the v.e.v.~with Robin boundary conditions as
\begin{equation}
    \langle \hat{\Phi}^2\rangle ^\zeta_0= \lim _{\rho '\to \rho }\frac{1}{4\pi^2}\sum_{\ell=-\infty}^{\infty}  \int_{-\infty}^{\infty} \, \left[g_{\omega\ell}^\zeta (\rho,\rho')-g_{\omega\ell}^N(\rho,\rho')\right]d\omega
    +\frac{\nu}{4\pi L} ,
    \label{eq:Robin_VP}
\end{equation}
where $g_{\omega\ell}^N(\rho,\rho')$ is the vacuum Euclidean Green's function (\ref{eq:gR}) with $\zeta =\pi/2$, corresponding to Neumann boundary conditions.

\begin{figure*}[htbp]
     \begin{center}
     \begin{tabular}{cc}
      \includegraphics[width=0.45\textwidth]{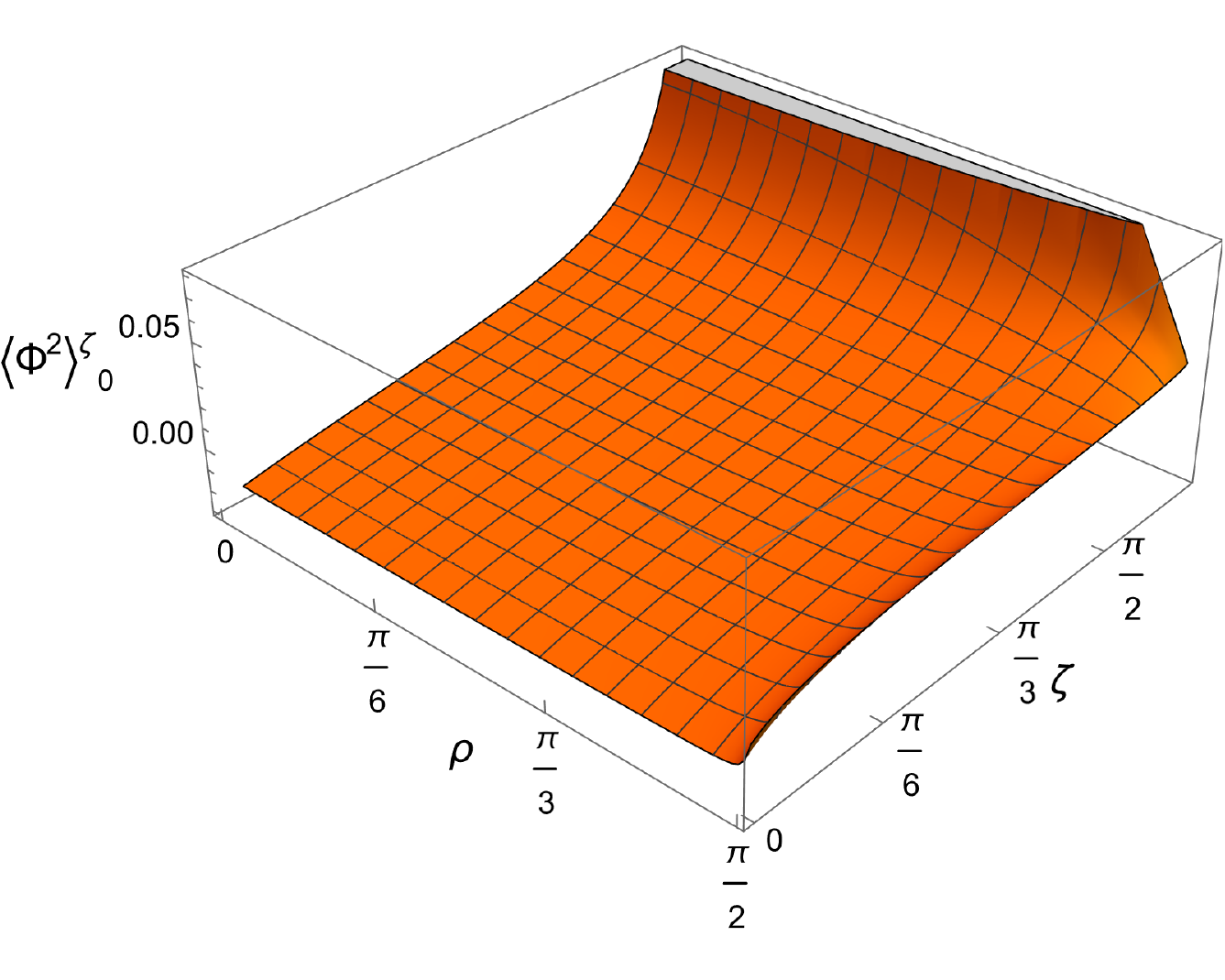} &
        \includegraphics[width=0.54\textwidth]{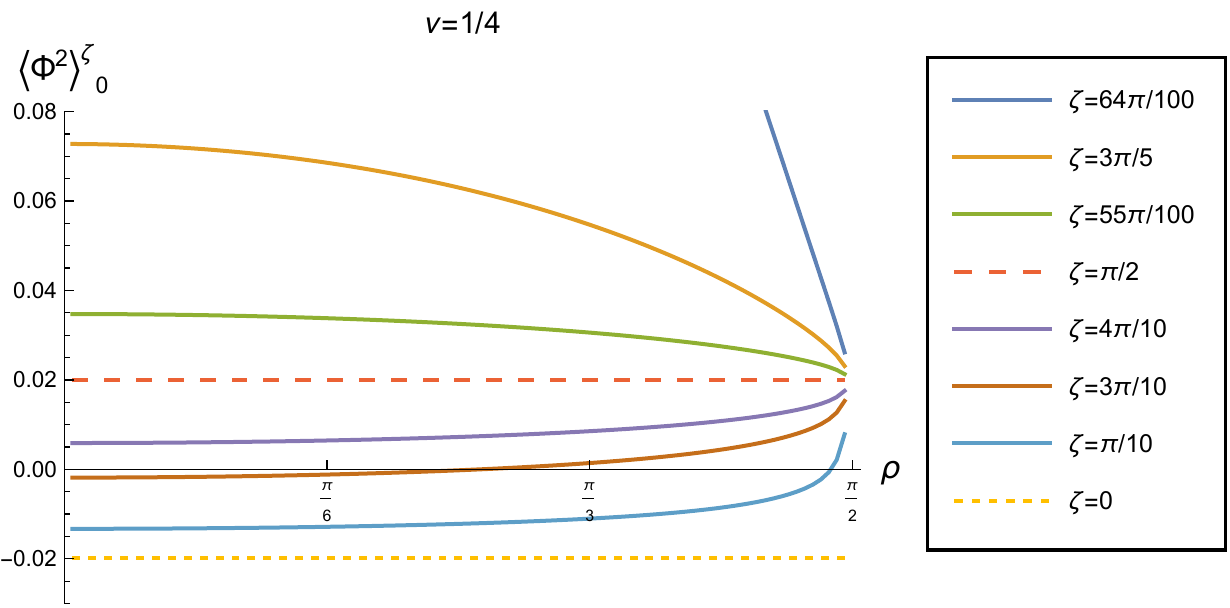}\\[0.3cm]
         \includegraphics[width=0.45\textwidth]{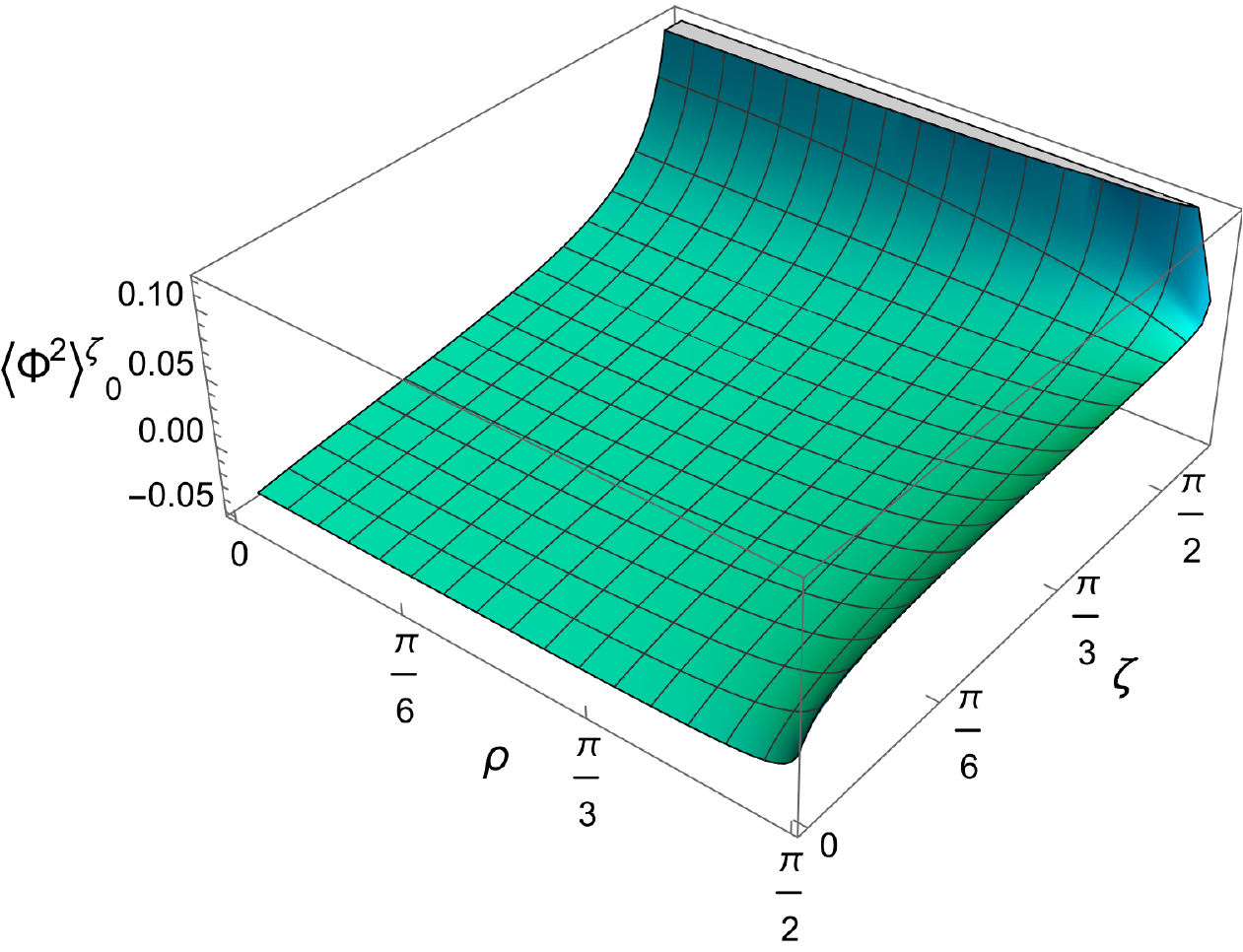} & 
         \includegraphics[width=0.54\textwidth]{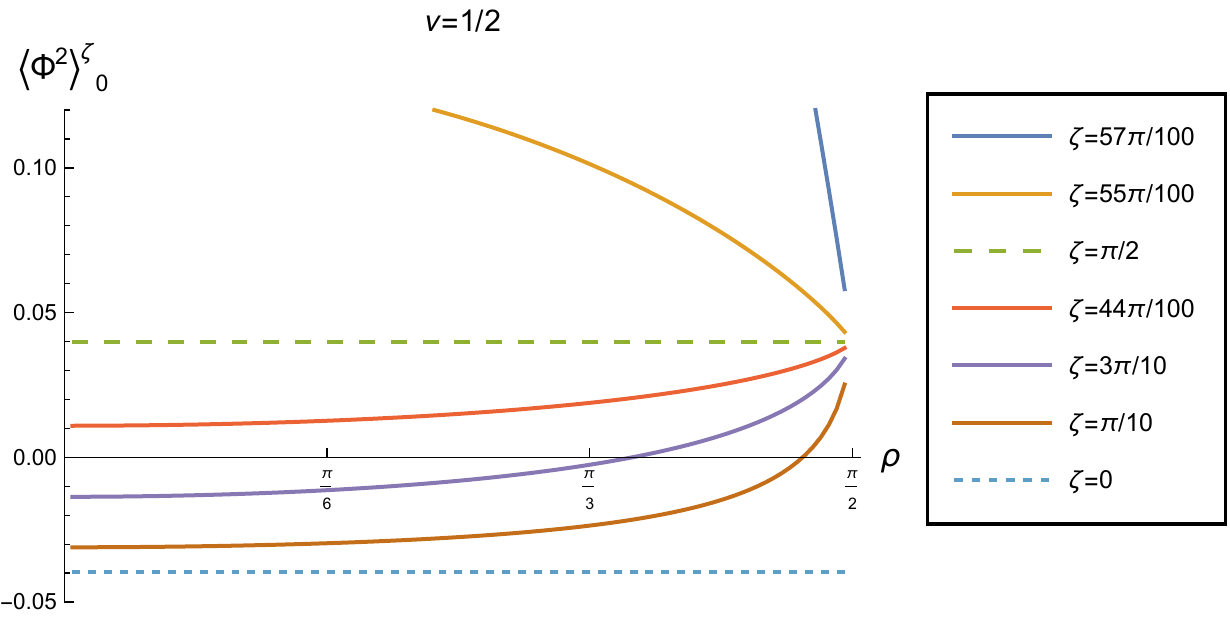}\\[0.3cm]
          \includegraphics[width=0.45\textwidth]{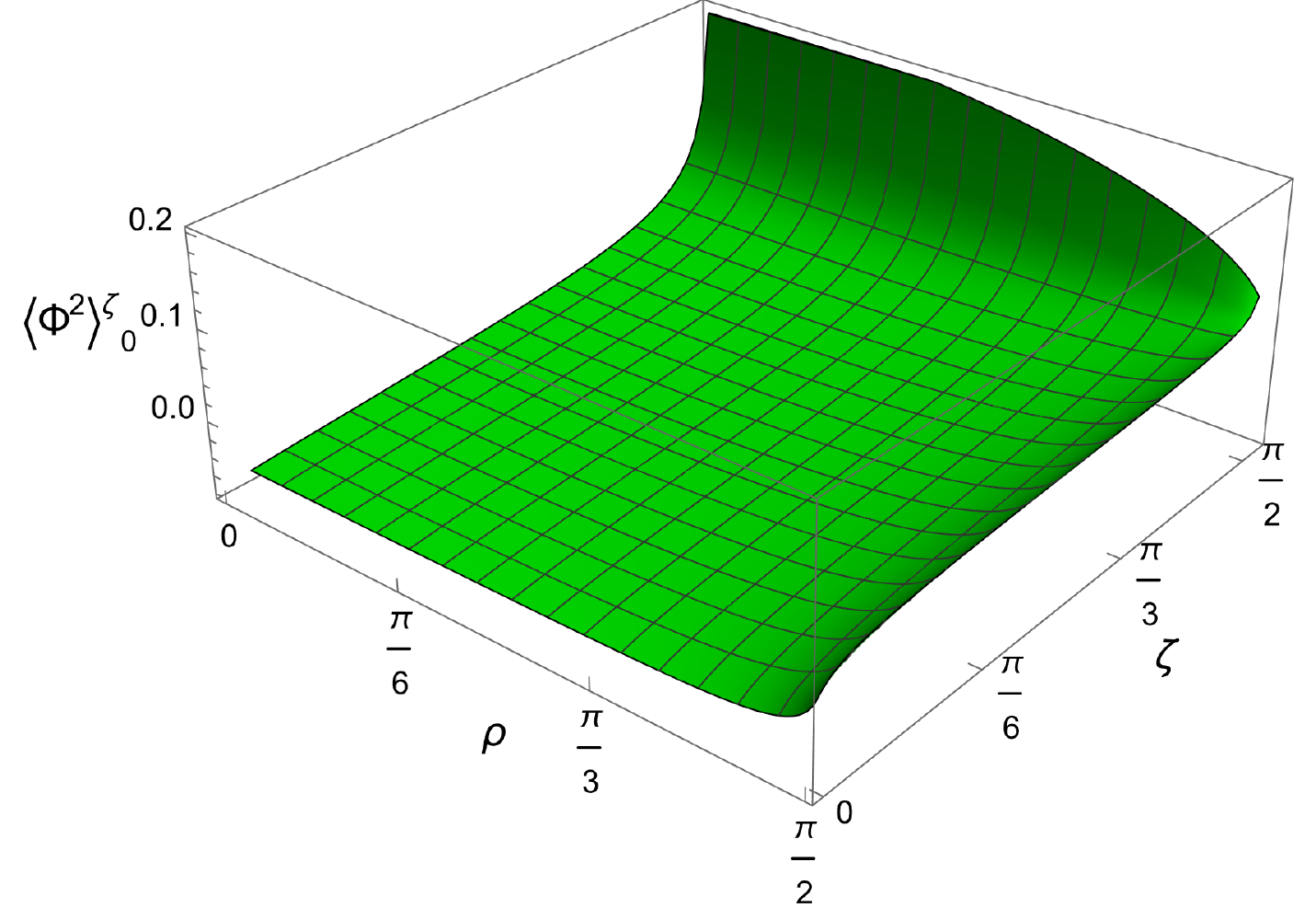} & 
          \includegraphics[width=0.54\textwidth]{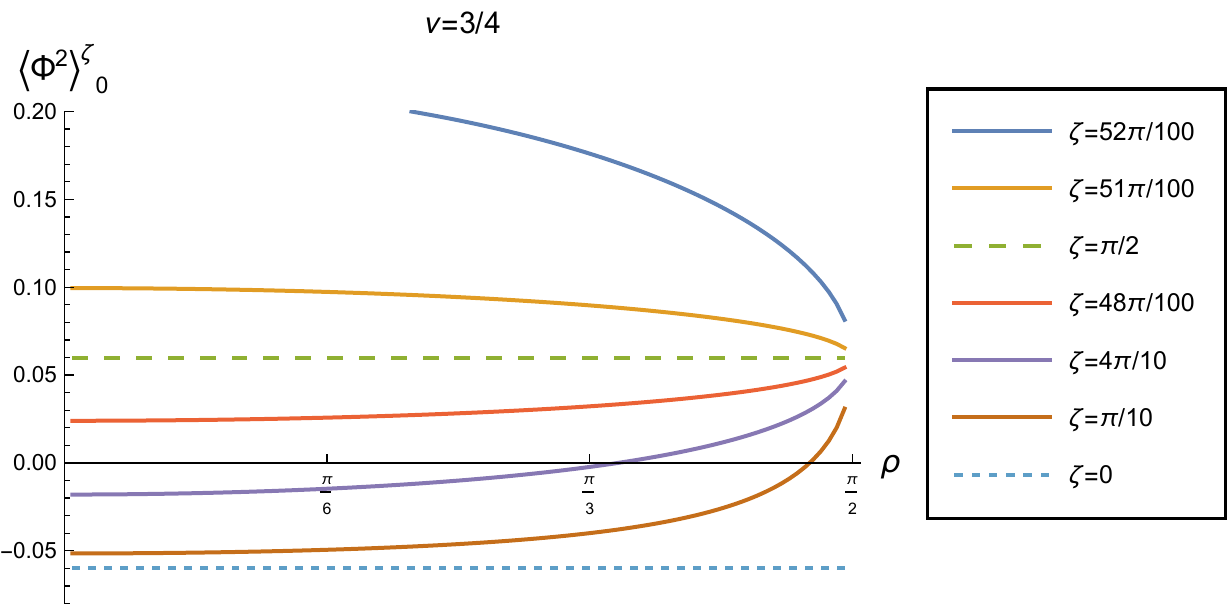} 
          \end{tabular}
               \caption{V.e.v.s of the VP, $\langle \hat{ \Phi}^2 \rangle _0^\zeta$,   with $\nu=1/4$ (top), $\nu =1/2$ (middle) and $\nu = 3/4$ (bottom). On the  left are the surface plots of $\langle \hat{ \Phi}^2 \rangle _0^\zeta$ as a function of $\rho$ and $\zeta$. On the right is shown $\langle \hat{ \Phi}^2 \rangle _0^\zeta$ as a function of $\rho$ for a selection of values of the Robin parameter $\zeta$. The dotted lines denote the results for Dirichlet ($\zeta=0$) and Neumann ($\zeta = \pi/2$) boundary conditions.}
        \label{fig:robin_vac_plots}
      \end{center}
\end{figure*}

We calculate the v.e.v.s (\ref{eq:Robin_VP}) with Robin boundary conditions for representative values of $\nu$, using {\footnotesize MATHEMATICA}. 
We find that the sum and integral in  \eqref{eq:Robin_VP} converge rapidly.  The results are  shown in figure~\ref{fig:robin_vac_plots}, where the v.e.v.s have been determined for different Robin parameters $0<\zeta <\zeta_{\text{crit}}$. 
As found for a massless, conformally coupled scalar field in four space-time dimensions \cite{Morley:2021}, the v.e.v.s with Robin boundary conditions are not constant in the space-time, unless we apply either Dirichlet or Neumann boundary conditions. 
For fixed $\nu $ and $\zeta $, the v.e.v.s are monotonic functions of $\rho $, but whether they are monotonically increasing or decreasing as $\rho $ increases depends on the value of $\zeta $. For all values of $\nu $ studied, the v.e.v.s are monotonically increasing as $\rho $ increases if $0<\zeta <\pi /2$, but monotonically decreasing for $\zeta >\pi /2$.
For each value of $\nu$ studied, the  v.e.v.s for fixed $\rho $ increase with increasing $\zeta$ and eventually diverge as $\zeta \to \zeta_{\text{crit}}$. With increasing $\nu$ and fixed $\rho $, $\zeta $, we see that the v.e.v.s all increase in magnitude for all $\rho$. Also it can be seen that the variation in the v.e.v.s is greatest at the origin of the space-time ($\rho=0$). As we reach the space-time boundary at $\rho=\pi/2$, the v.e.v.s for all Robin parameters (except the Dirichlet case $\zeta=0$) converge to the result for Neumann boundary conditions. 
This generalizes the finding of \cite{Morley:2021} in the massless, conformally coupled case in four dimensions. 
   
\subsection{Thermal expectation values}
\label{subsec:Robin_thermal}
\begin{figure}[htbp]
	\begin{center}
		\begin{tabular}{cc}
			\includegraphics[width=0.45\textwidth]{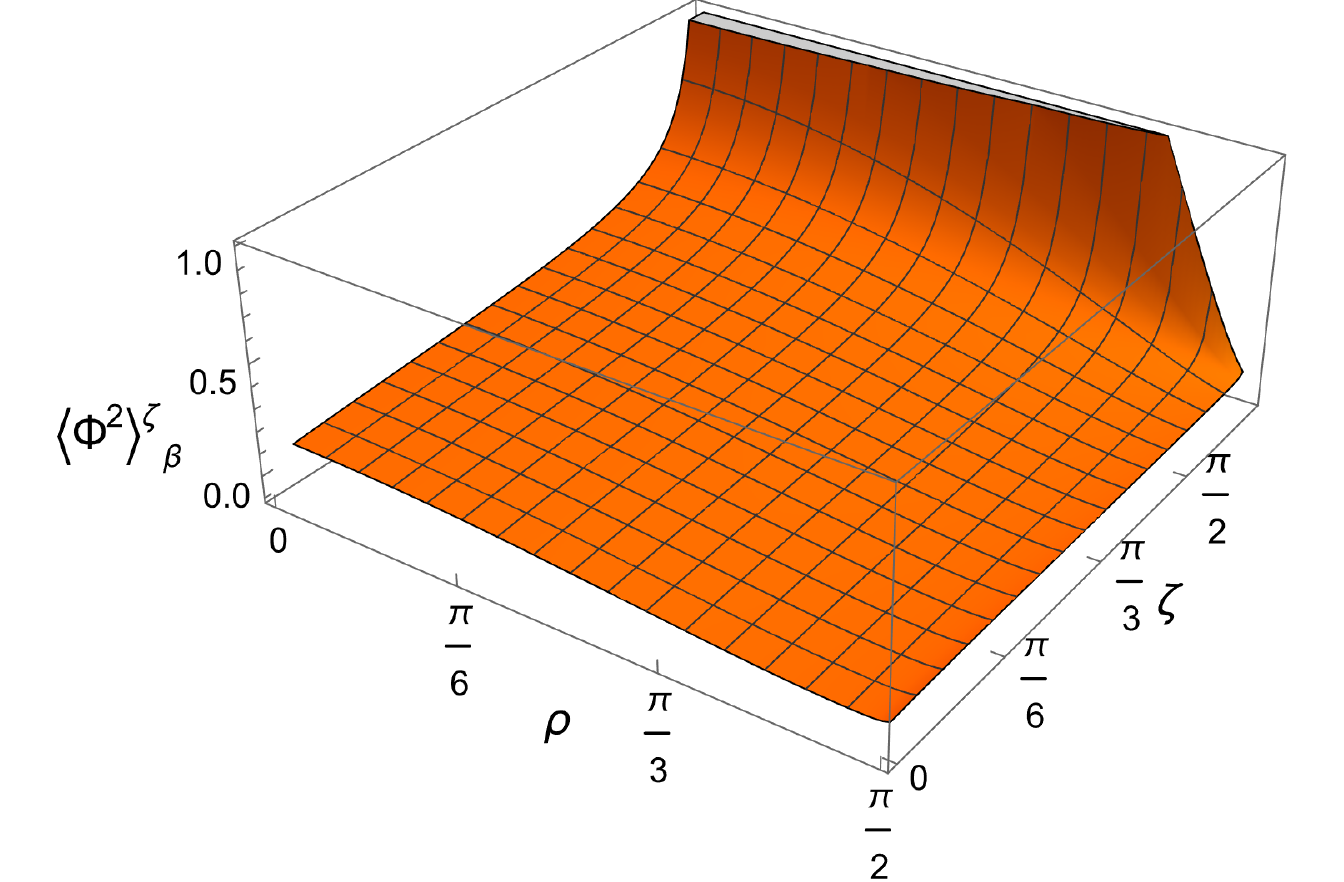} & 
			\includegraphics[width=0.54\textwidth]{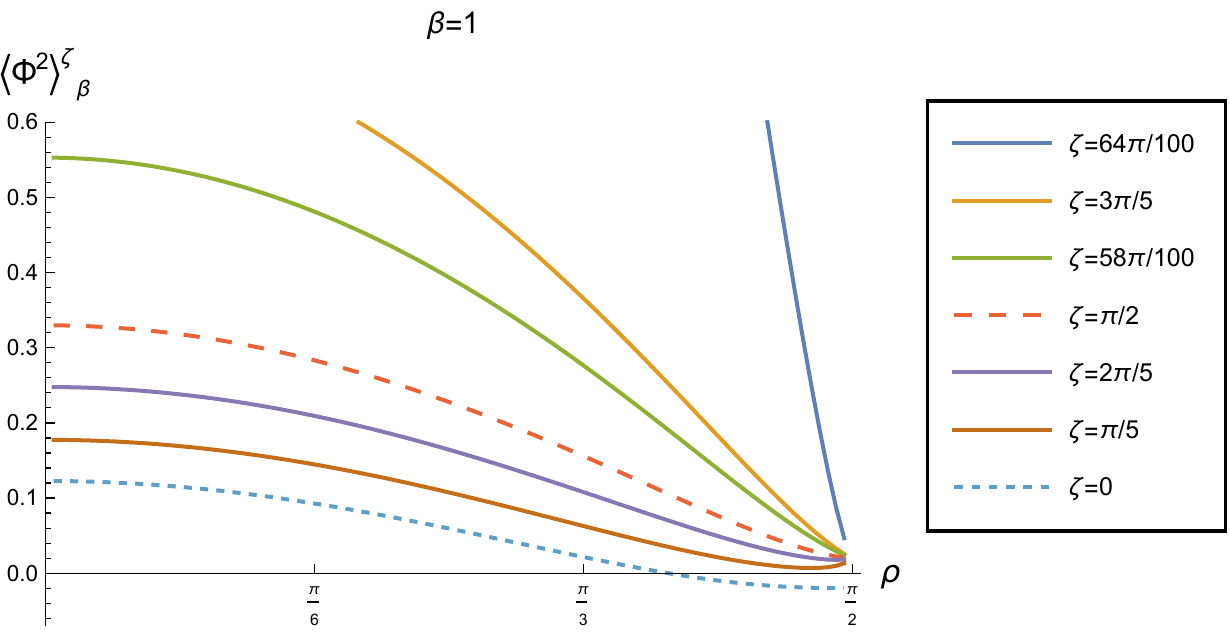}
			\\[0.3cm]
			\includegraphics[width=0.45\textwidth]{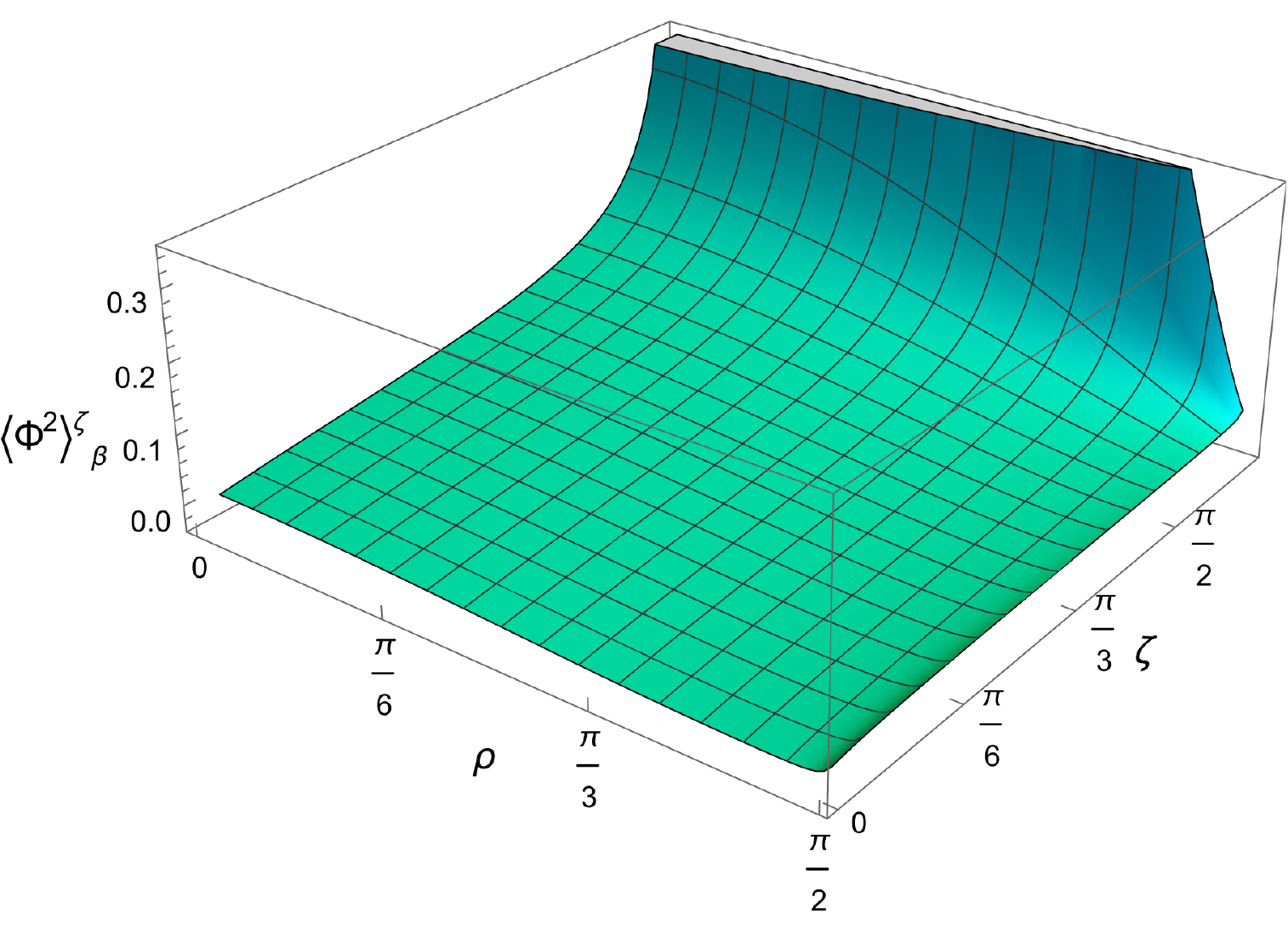} &
			\includegraphics[width=0.54\textwidth]{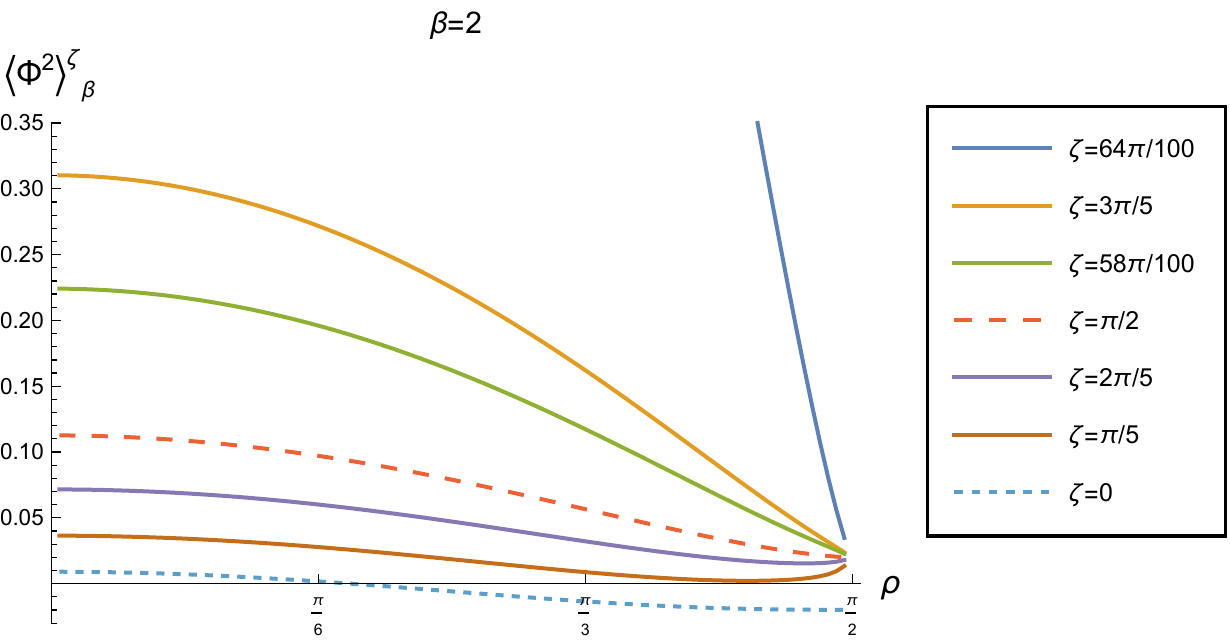}
			\\[0.3cm]
			\includegraphics[width=0.45\textwidth]{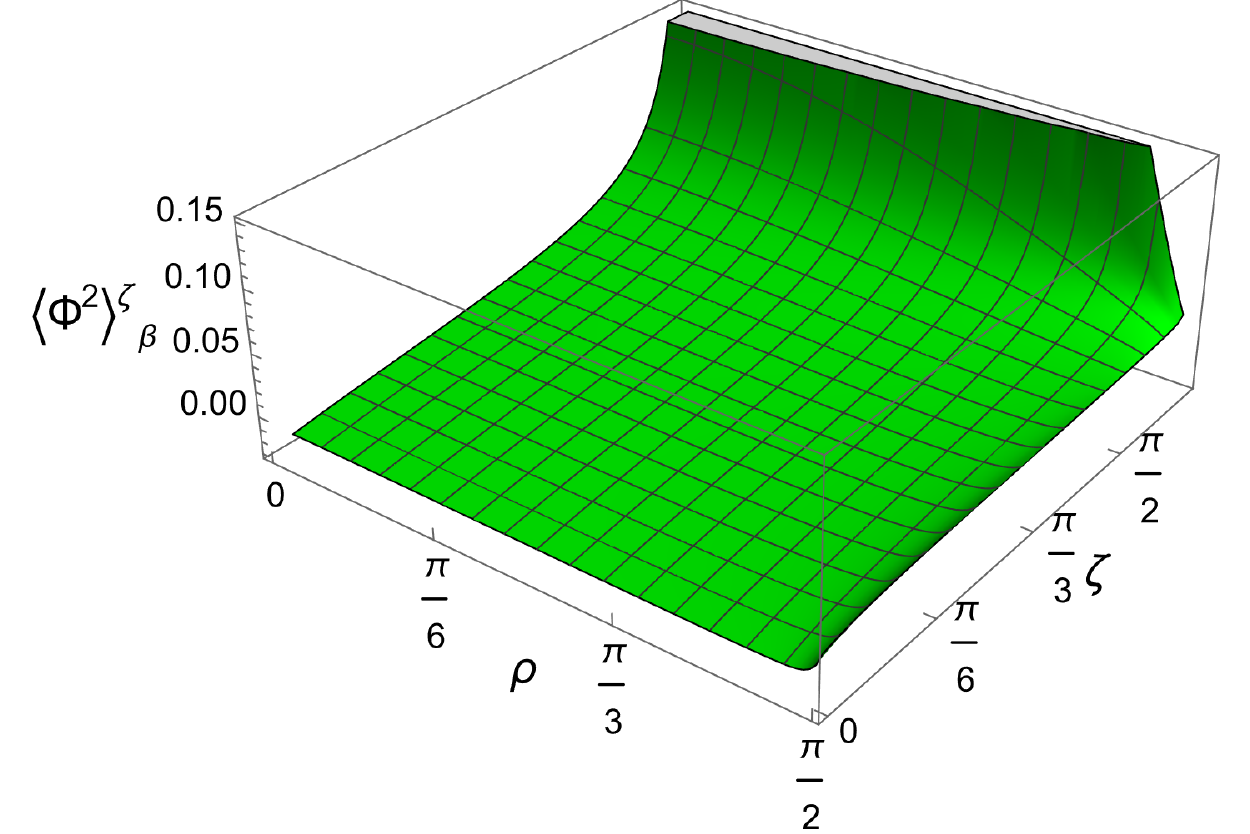} &
			\includegraphics[width=0.54\textwidth]{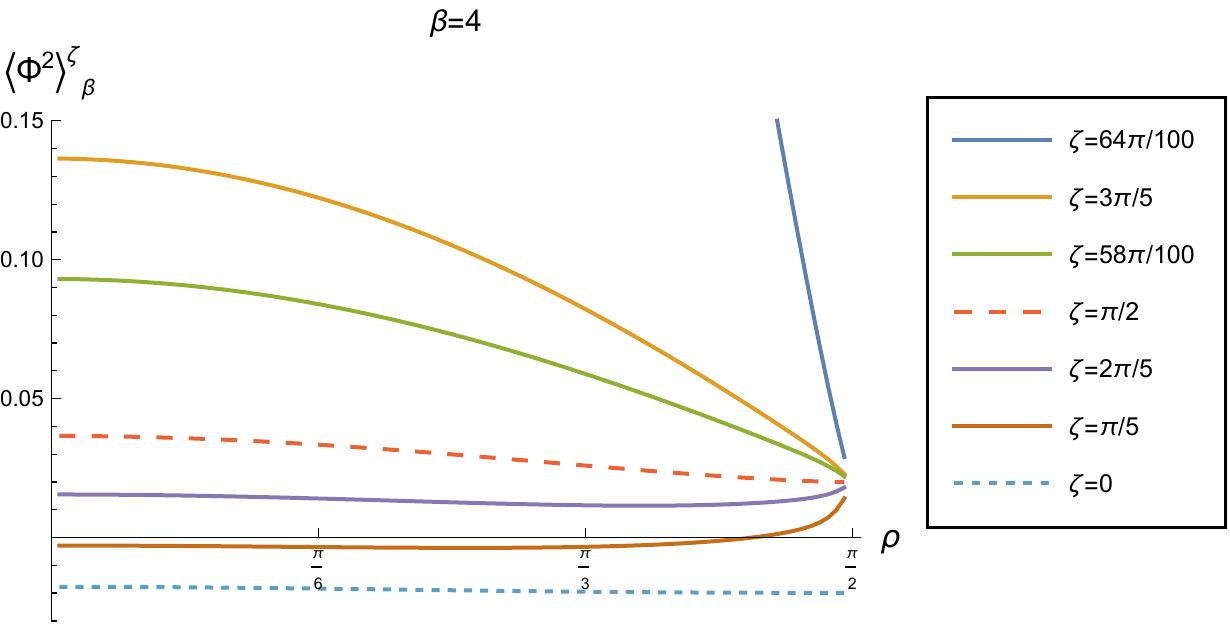}
		\end{tabular}
		\caption{T.e.v.s of the VP, $\langle \hat{ \Phi}^2 \rangle _\beta^\zeta$, for  $\nu=1/4$ and a selection of values of the inverse temperature $\beta$: $\beta =1$ (top), $\beta = 2$ (middle), $\beta = 4$ (bottom). 
			On the left are the 3D surface plots of $\langle \hat{ \Phi}^2 \rangle _\beta^\zeta$ as a  function of $\rho$ and $\zeta$.
			On the right is shown $\langle \hat{ \Phi}^2 \rangle _\beta^\zeta$ as a function of $\rho$ for  various values of $\zeta$. The dotted lines denote the results for Dirichlet and Neumann boundary conditions. }
		\label{fig:multiple}
	\end{center}
\end{figure}

\begin{figure}[htbp]
	\begin{center}
		\begin{tabular}{cc}
			\includegraphics[width=0.45\textwidth]{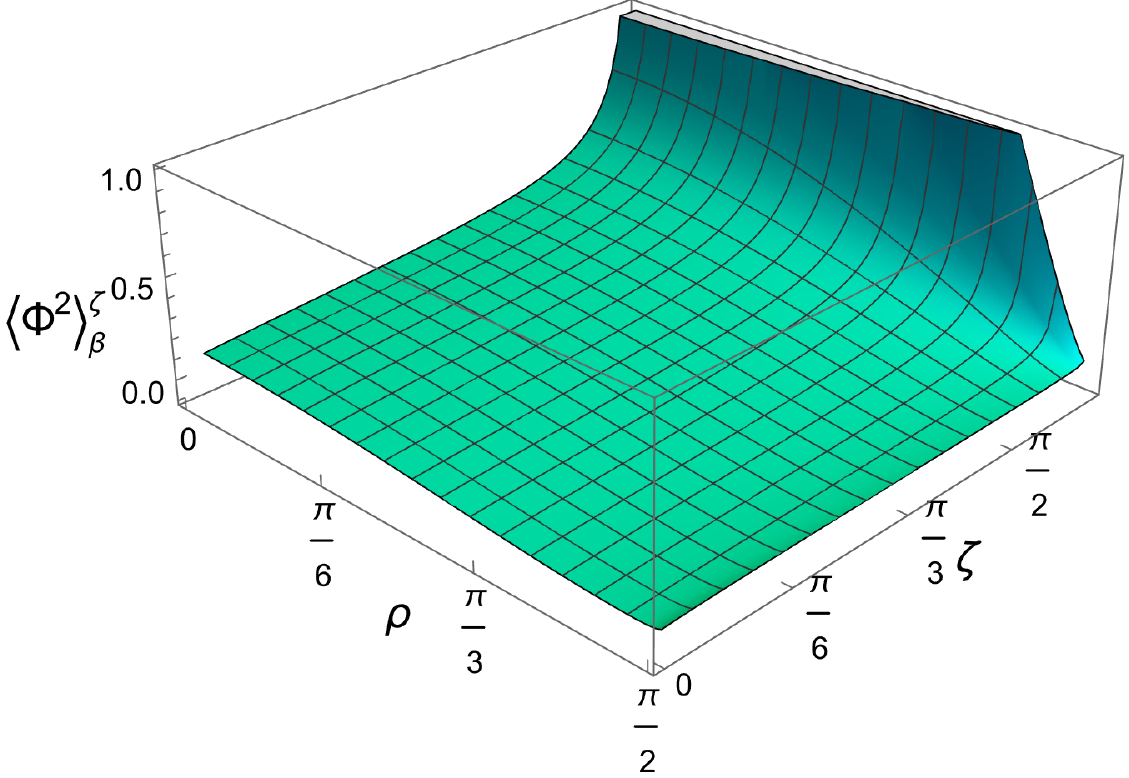} &
			\includegraphics[width=0.54\textwidth]{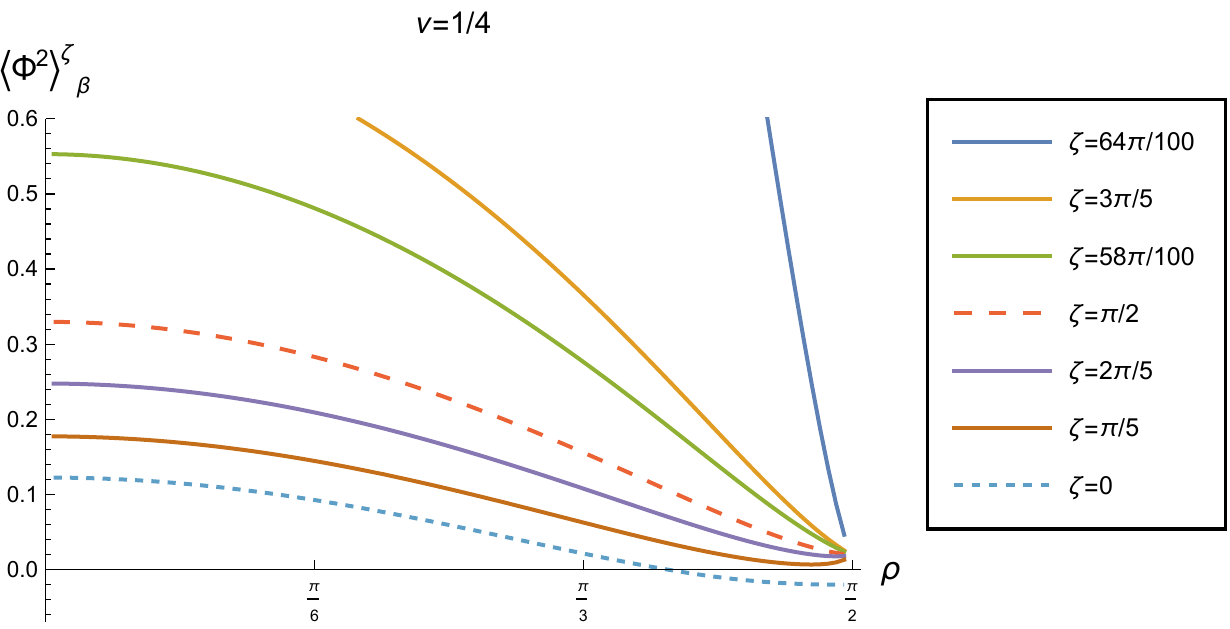}         \\[0.3cm]
			\includegraphics[width=0.45\textwidth]{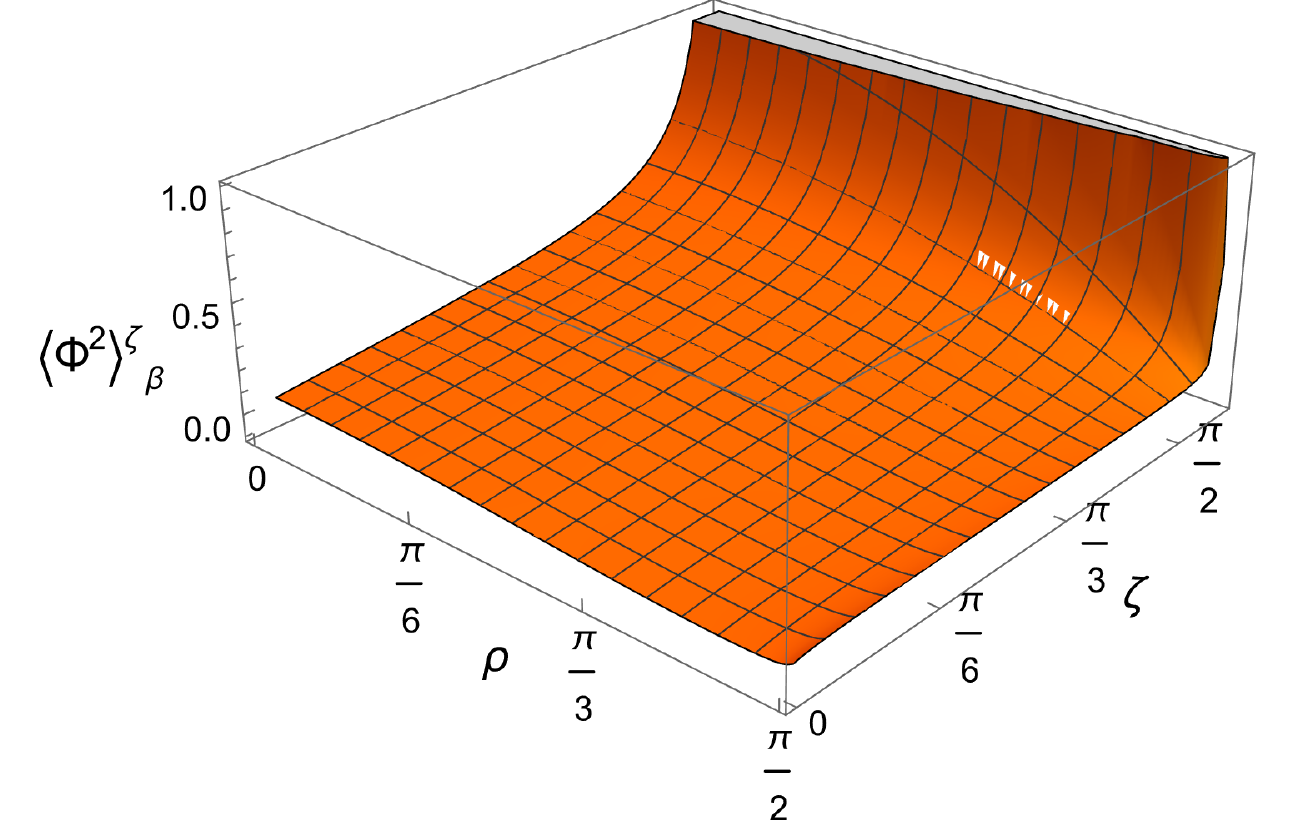} & 
			\includegraphics[width=0.54\textwidth]{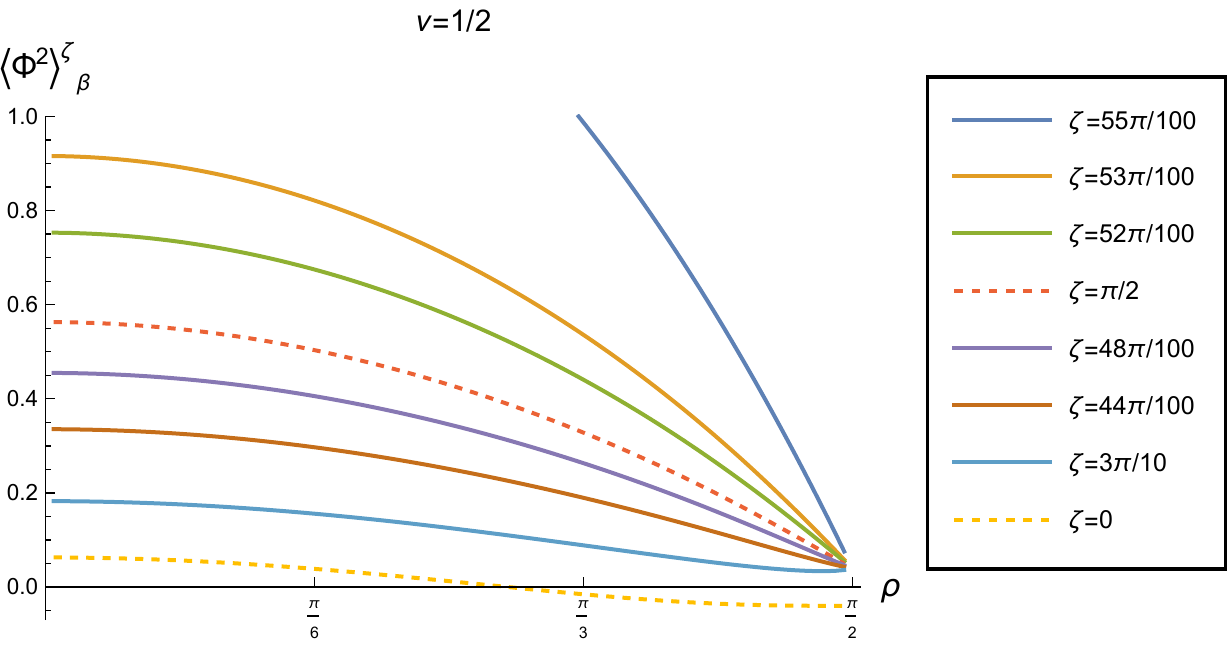}\\[0.3cm]
			\includegraphics[width=0.45\textwidth]{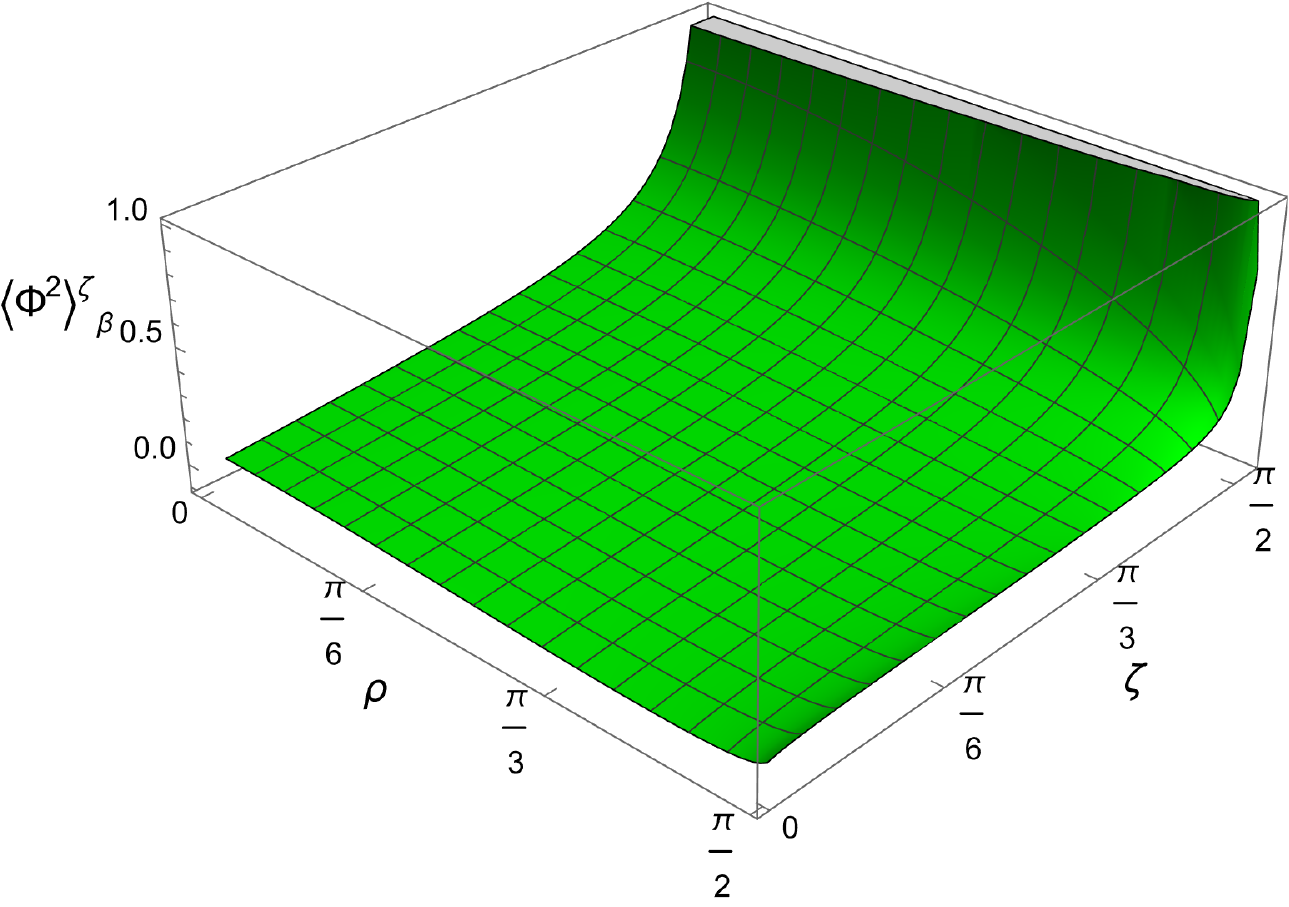} &
			\includegraphics[width=0.54\textwidth]{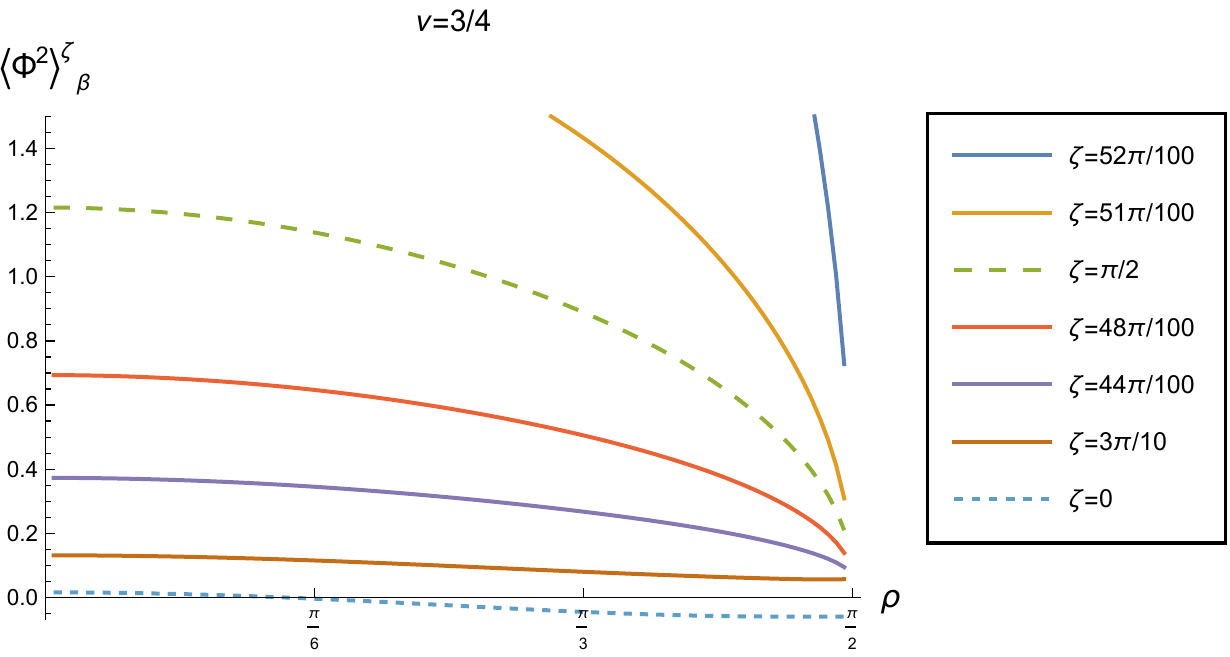}
		\end{tabular}
		\caption{T.e.v.s of the VP, $\langle \hat{ \Phi}^2 \rangle _0^\zeta$, with inverse temperature $\beta = 1$,  for  a selection of values of $\nu$: $\nu =1/4$ (top), $\nu = 1/2$ (middle), $\nu = 3/4$ (bottom). On the left is shown the 3D surface plots of $\langle \hat{ \Phi}^2 \rangle _0^\zeta$ as a function of $\rho$ and $\zeta$. On the right is shown $\langle \hat{ \Phi}^2 \rangle _0^\zeta$ as a function of $\rho$ for various values of the Robin parameter $\zeta$, with the dotted lines denoting the results for Dirichlet and Neumann boundary conditions.}
		\label{fig:robin_thermal}
	\end{center}
\end{figure}

For the thermal Euclidean Green's functions (\ref{eq:Euclid_Thermal_Green}) we replace $\omega$ in (\ref{eq:gR}) with $n \kappa$ for $n \in \mathbb{Z}$ and $\kappa$ given in \eqref{eq:kappa}. To calculate the t.e.v.s, we employ the same method 
used in section~\ref{subsec:Robin_vac} and consider  the difference between the unrenormalised t.e.v.s with Robin and Neumann boundary conditions. Following this we  add the t.e.v.~with Neumann boundary conditions previously determined in (\ref{eq:VPN}, \ref{eq:thermalN}). We can therefore write 
\begin{multline}
    \langle \hat{\Phi}^2\rangle ^\zeta_\beta =  \frac{1}{2\pi\beta}\lim _{\rho' \to \rho }\sum_{\ell=-\infty}^{\infty}\,\,  \sum_{n=-\infty}^{\infty} \, \left[g_{n\ell }^\zeta (\rho,\rho')-g_{n\ell }^N(\rho,\rho')\right]
    +\frac{\nu}{4\pi L} \\
    +\frac{1}{2 \pi L} \sum_{j=1}^{\infty}  \left\{ \nu F\left (1+ \nu,1-\nu,\frac{3}{2};\frac{1-\cosh(j\beta)}{2\cos^2\rho}\right) 
\right. \\ \left. 
+ \frac{\sqrt{2}\cos\rho}{2 \sqrt{\cosh(j\beta)-1}} F \left(\frac{1}{2}+\nu,\frac{1}{2}-\nu,\frac{1}{2};\frac{1-\cosh(j\beta)}{2\cos^2\rho}\right) \right\}.
    \label{eq:Robin_thermal_VP}
\end{multline}
We calculate the t.e.v.s of the VP, with Robin boundary conditions, using {\footnotesize MATHEMATICA} for different values of $\nu$ and Robin parameter~$\zeta < \zeta_{\text{crit}}$. Our results are displayed in figures~\ref{fig:multiple} and \ref{fig:robin_thermal}.

Figure~\ref{fig:multiple} shows the t.e.v.s for different inverse temperature, $ \beta$, with $\nu=1/4$. As with the v.e.v.s, the t.e.v.s for fixed $\rho $ increase with increasing $\zeta$ and start to diverge as  $\zeta_{\text{crit}}$ is approached. We also  see that with increasing $\beta$, and thus decreasing temperature, the plots resemble the corresponding v.e.v.s, a finding also noted in \cite{Morley:2021}.  Figure~\ref{fig:robin_thermal} shows the t.e.v.s for three different values of $\nu$ and inverse temperature $\beta=1$. As $\nu$ increases, we see the values of the t.e.v.s increasing in magnitude for all $\rho$, as in the vacuum case. Both figures~\ref{fig:multiple} and \ref{fig:robin_thermal} show that the maximum difference between the t.e.v.s, for the different Robin parameters, is found at the space-time origin, as was found for the t.e.v.s with Dirichlet and Neumann boundary conditions (see figure~\ref{fig:thermal}). As we reach the space-time boundary, all t.e.v.s converge to the Neumann v.e.v.s, except for the Dirichlet case which converges to its own v.e.v.. 
For a massless, conformally coupled scalar field in four dimensions, similar results were obtained in 
\cite{Morley:2021}.  
Our work therefore shows that the key finding in \cite{Morley:2021}, namely that near the boundary the Neumann v.e.v.s are generic and those for  Dirichlet boundary conditions a special case, extends to three dimensions and remains true for all scalar field masses and values of the coupling constant for which Robin boundary conditions can be applied.

\section{Vacuum polarisation at the space-time boundary}
\label{sec:boundary}
In sections~\ref{subsec:Robin_vac}, \ref{subsec:Robin_thermal} we found that the VP with Robin boundary conditions  converges to the Neumann v.e.v.~at the space-time boundary for all values of the Robin parameter $0<\zeta <\zeta_{\text{crit}}$ and for all values of the scalar field mass and coupling.
In \cite{Morley:2021}, this result was demonstrated using an analytic argument for a massless and conformally coupled scalar field in four dimensions. 
In this section we extend the argument of \cite{Morley:2021} to a massless, conformally coupled scalar field ($\nu = 1/2$) on three-dimensional adS, before examining more general mass and coupling.
The analysis for $\nu =1/2$ very closely follows that in \cite{Morley:2021}, so our presentation is brief; further details can be found in \cite{Deutsch:1978sc, Morley:2021}.

For a massless, conformally coupled scalar field, the behaviour of the VP at the boundary can be studied by transforming to ESU and applying the general methodology of \cite{Deutsch:1978sc}. This enables us to write $G^{\rm {ESU}}_{\zeta }(x,x')$, the vacuum Euclidean Green's function on ESU with Robin boundary conditions,  as an asymptotic series:
\begin{align}
    G^{\rm {ESU}}_{\zeta }(x,x') & = G^{\rm {ESU}}_{{N}}(x,x') - \frac{\cot \zeta }{L}
    \int _{{{\mathcal {I}}}_{\frac{\pi }{2}}} G^{\rm {ESU}}_{{N}}(x,y)G^{\rm {ESU}}_{\zeta }(y,x') \, dS
    \nonumber \\ 
    & =  G^{\rm {ESU}}_{{N}}(x,x') - \frac{\cot \zeta }{L}
    \int _{{{\mathcal {I}}}_{\frac{\pi }{2}}} G^{\rm {ESU}}_{{N}}(x,y)G^{\rm {ESU}}_{{N}}(y,x') \, dS 
    \nonumber \\ & \qquad \qquad
    +
     \frac{\cot ^{2}\zeta }{L^{2}}  \int _{{{\mathcal {I}}}_{\frac{\pi }{2}}} G^{\rm {ESU}}_{{N}}(x,y) \left[  \int _{{{\mathcal {I}}}_{\frac{\pi }{2}}} G^{\rm {ESU}}_{{N}}(y,z)G^{\rm {ESU}}_{{N}}(z,x') \, dS  \right] \, dS
    \ldots ,
    \label{eq:CD}
\end{align}
where $G^{\rm {ESU}}_{{N}}(x,x')$ is the vacuum Euclidean Green's function on ESU with Neumann boundary conditions applied:
\begin{multline}
   G^{\rm {ESU}}_{{N}}(x,x') = \frac{1}{4{\sqrt {2}}\pi L} \left\{  
   \left[ \cosh \Delta \tau  - \cos \Delta \theta  \sin \rho \sin \rho ' + \cos \rho \cos \rho '  \right] ^{-\frac{1}{2}}
   \right. \\ \left. 
   + \left[ 
  \cosh \Delta \tau  - \cos \Delta \theta \sin \rho \sin \rho '  - \cos \rho \cos \rho '\right] ^{-\frac{1}{2}}
   \right\} .
\end{multline}
The first line in (\ref{eq:CD}) arises from applying Stokes' theorem in a region of ESU bounded by timelike hypersurfaces ${\mathcal {I}}_{0}$ at $\rho =0$ and ${\mathcal{I}}_{\frac{\pi }{2}}$ at $\rho = \pi/2$ (the surface integral over ${\mathcal {I}}_{0}$ vanishes via the same argument as in \cite{Morley:2021}).
In the second line of (\ref{eq:CD}) we use an iterative method, replacing $G^{\rm {ESU}}_{\zeta }(x,x')$ in the integral by the expression on the right-hand-side of the first line and successively repeating this step.  
The integrals in (\ref{eq:CD}) are over the space-time points $y$ and $z$.

We transform back to adS by multiplying (\ref{eq:CD}) throughout by ${\sqrt{ \cos \rho \cos \rho'}}$. 
Then, taking the limit $x' \to x$, we have the following relationship between the VP on adS with Robin boundary conditions applied and that with Neumann boundary conditions 
\begin{multline}
    \langle {\hat {\Phi }}^{2} \rangle _{0}^{\zeta } =\langle {\hat {\Phi }}^{2} \rangle _{0}^{{{N}}} 
     - \frac{\cos \rho \cot \zeta }{L}
    \int _{{{\mathcal {I}}}_{\frac{\pi }{2}}} G^{\rm {ESU}}_{{N}}(x,y)G^{\rm {ESU}}_{{N}}(y,x) \, dS 
    \\
    +
     \frac{\cos \rho \cot ^{2}\zeta }{L^{2}}  \int _{{{\mathcal {I}}}_{\frac{\pi }{2}}} G^{\rm {ESU}}_{{N}}(x,y) \left[  \int _{{{\mathcal {I}}}_{\frac{\pi }{2}}} G^{\rm {ESU}}_{{N}}(y,z)G^{\rm {ESU}}_{{N}}(z,x) \, dS  \right] \, dS
    \ldots ,
    \label{eq:CD1}
\end{multline}
 which is valid for all nonzero $\zeta $, that is, for all boundary conditions other than Dirichlet.
Next we evaluate the first integral on the right-hand-side of (\ref{eq:CD1}):
\begin{align}
   \int _{{{\mathcal {I}}}_{\frac{\pi }{2}}} G^{\rm {ESU}}_{{N}}(x,y)G^{\rm {ESU}}_{{N}}(y,x) \, dS 
   & = 
  \frac{1}{8\pi ^{2}L^{2}} \int _{{{\mathcal {I}}}_{\frac{\pi }{2}}} 
  \frac{1}{\cosh \Delta \tau - \cos \Delta \theta \sin \rho } \, dS
  \nonumber \\ & = \frac{1}{4\pi } \int _{\Delta \tau =-\infty }^{\infty } \frac{1}{{\sqrt {\cosh ^{2}\Delta \tau - \sin ^{2}\rho }}} \, d\Delta \tau 
  \nonumber \\ & = 
  \frac{1}{2\pi } K(-\tan ^{2}\rho ) \sec \rho ,
  \label{eq:CD2}
\end{align}
where $K$ is the complete elliptic integral of the first kind \cite[19.2.8]{NIST:DLMF}. 
Here $x$ is a general point in ESU with coordinates $(\tau , \rho, \theta)$, the integral is performed over points $y=(\tau _{y},\pi/2,\theta _{y})$ on ${\mathcal {I}}_{\frac{\pi }{2}}$ and $\Delta \tau = \tau _{y}-\tau $, $\Delta \theta = \theta _{y}-\theta $.
In the second line in (\ref{eq:CD2}) we have performed the integral over $\Delta \theta \in [0,2\pi ]$ and in the third line the integral over $\Delta \tau $ is performed. 
As the point $x$ approaches the boundary at ${\mathcal {I}}_{\frac{\pi }{2}}$, we have
\begin{equation}
    \int _{{{\mathcal {I}}}_{\frac{\pi }{2}}} G^{\rm {ESU}}_{{N}}(x,y)G^{\rm {ESU}}_{{N}}(y,x) \, dS  \sim \frac{1}{2\pi } \log \epsilon 
\end{equation}
for $\rho = \pi /2-\epsilon $ as $\epsilon \to 0 $.
This diverges as the boundary is approached, but the divergence is weaker than that of the corresponding integral in four dimensions \cite{Morley:2021}. 
The second integral in (\ref{eq:CD1}) remains finite as the point $x$ approaches ${\mathcal {I}}_{\frac{\pi }{2}}$ and does not need to be considered further. 
Substituting in (\ref{eq:CD1}) then gives
\begin{equation}
    \langle {\hat {\Phi }}^{2} \rangle _{0}^{\zeta } =\langle {\hat {\Phi }}^{2} \rangle _{0}^{N}  -\frac{ \cot \zeta}{2 \pi L}\, K(-\tan^2\rho) + \ldots ,
    \label{eq:Kint2}
\end{equation}
where $\ldots $ denotes terms which vanish as $\rho \to \pi/2$. In this limit the elliptic integral also vanishes and hence 
\begin{equation}
    \lim _{\rho \to \frac{\pi }{2}} \langle {\hat {\Phi }}^{2} \rangle _{0}^{\zeta } =\lim _{\rho \to \frac{\pi }{2}} \langle {\hat {\Phi }}^{2} \rangle _{0}^{N} .
\end{equation}
Here we have considered v.e.v.s, but the argument extends trivially to t.e.v.s.  
However, the above analysis does depend crucially on a conformal transformation to ESU, and hence can only be applied when $\nu =1/2$. For other values of $\nu $, we take an alternative approach.

Consider first the v.e.v.s, determined using \eqref{eq:Robin_VP}. The first part of the right-hand-side of \eqref{eq:Robin_VP} is the difference between v.e.v.s with Robin and Neumann boundary conditions applied. We can write this as 
\begin{equation}
    \langle {\hat {\Phi }}^{2} \rangle _{0}^{\zeta } - \langle {\hat {\Phi }}^{2} \rangle _{0}^{N}
    = \frac{1}{4\pi^2}\sum_{\ell=-\infty}^{\infty}  \int_{-\infty}^{\infty} \, p_{\omega \ell}(\rho)\left[\mathcal{N}_{\omega \ell}^\zeta \,q_{\omega\ell}^\zeta (\rho)- \mathcal{N}_{\omega \ell}^N\, q_{\omega\ell}^N(\rho)\right]d\omega
    \label{eq:VP_boundary1}
\end{equation}
where we have used~(\ref{eq:g0}, \ref{eq:g1}, \ref{eq:greencon}) and, as before, the $\zeta, N$ superscripts refer to the Robin and Neumann boundary conditions respectively. 
Using (\ref{eq:gR}), we can express \eqref{eq:VP_boundary1} as 
\begin{equation}
   \langle {\hat {\Phi }}^{2} \rangle _{0}^{\zeta } - \langle {\hat {\Phi }}^{2} \rangle _{0}^{N} = -\frac{L\nu \cos \zeta}{2\pi^2} \sum_{\ell=-\infty}^{\infty} \int^{\infty}_{-\infty}  \mathcal{N}_{\omega \ell}^\zeta \,\, \mathcal{N}_{\omega \ell}^N \, \,[p_{\omega \ell} (\rho)]^2 \, d\omega . 
   \label{eq:VPB2}
\end{equation}
To analyze the behaviour of this quantity as $\rho \to \pi/2$, it is helpful to use the form (\ref{genrad}) for the radial function $p_{\omega \ell }(\rho )$. 
As the space-time boundary is approached, the hypergeometric functions (\ref{genrad}) tend to unity. Therefore the second term in (\ref{genrad}) is dominant, giving the following leading-order behaviour of \eqref{eq:VPB2}: 
\begin{equation}
    \langle {\hat {\Phi }}^{2} \rangle _{0}^{\zeta } - \langle {\hat {\Phi }}^{2} \rangle _{0}^{N}  \sim  -\frac{L\nu [\cos \rho]^{2-2\nu} \cos \zeta}{2\pi^2}\sum_{\ell=-\infty}^{\infty}
   \int^{\infty}_{-\infty}  \mathcal{N}_{\omega \ell}^\zeta \,\, \mathcal{N}_{\omega \ell}^N \,  [\sin \rho]^{2 \lvert \ell  \rvert }\mathcal{Q}^2 \, d\omega ,
   \label{eq:VPB3}
\end{equation}
where $\mathcal{Q}$ is given in \eqref{eq:pandq}.

The integral over $\omega $ and sum over $\ell $  in \eqref{eq:VPB3} cannot be performed analytically. 
However, whether or not this quantity is convergent (and the rate at which it diverges if it is divergent) depends only on the behaviour for large $\lvert \ell \rvert $, $\lvert \omega \rvert $.
Using the expression (\ref{eq:NnlR}) for the normalization constants, and considering only the dominant behaviour of $\mathcal{N}_{\omega \ell}^\zeta\mathcal{N}_{\omega \ell}^N \,  \mathcal{Q}^2$ for large $\lvert \ell \rvert $, $\lvert \omega \rvert $ gives
\begin{equation}
\mathcal{N}_{\omega \ell}^\zeta\mathcal{N}_{\omega \ell}^N \, \mathcal{Q}^2 \sim \frac{1}{4L^{2}\nu ^{2}}
\frac{\Gamma (\nu )^{2}}{\Gamma (-v)^{2}} 
 \frac{\lvert \Gamma(1/2-\nu/2+\lvert \ell \rvert /2 +i\omega/2)\rvert ^4}{\lvert \Gamma(1/2 + \nu/2 +\lvert \ell \rvert /2 + i \omega/2)\rvert ^4}
\csc \zeta  .
\end{equation}
Clearly this expression is valid only for $\zeta >0$, that is for all boundary conditions except Dirichlet boundary conditions.
Let us therefore consider the quantity
\begin{equation}
    \Sigma _{0} =\sum_{\ell=-\infty}^{\infty}  \int^{\infty}_{-\infty}  \frac{\lvert \Gamma(1/2-\nu/2+\lvert \ell \rvert /2 +i\omega/2)\rvert ^4}{\lvert \Gamma(1/2 + \nu/2 +\lvert \ell \rvert /2 + i \omega/2)\rvert ^4} [\sin \rho]^{2\lvert \ell \rvert } \, d\omega .
    \label{eq:VPB4}
\end{equation}
We are still unable to perform the sum over $\ell $ and integral over $\omega $ in \eqref{eq:VPB4} analytically. However, from \cite[5.11.12]{NIST:DLMF}, for large $\lvert \ell \rvert  $ and $\lvert \omega  \rvert $ we have
\begin{equation}
   \frac{\lvert \Gamma(1/2-\nu/2+\lvert \ell \rvert /2 +i\omega/2)\rvert ^4}{\lvert \Gamma(1/2 + \nu/2 +\lvert \ell \rvert /2 + i \omega/2)\rvert ^4} \sim \frac{2^{4\nu }}{(\lvert \ell \rvert ^2 + \lvert \omega \rvert ^2)^{2\nu}} .
\end{equation}
Using this approximation, the expression
\begin{equation}
    \Sigma _{1} = \sum_{\ell =1}^\infty \int_{0}^\infty  \frac{2^{4\nu }[\sin\rho] ^{2\ell}}{(\ell^2 + \omega ^2)^{2\nu}} d \omega 
   \label{eq:VPB5} 
\end{equation}
is amenable to exact evaluation.
The integral over $\omega $ in \eqref{eq:VPB5} is convergent for $\nu > 1/4$, leading to 
\begin{equation}
    \Sigma _{1} = \frac{2^{4\nu }\sqrt{\pi}\,\nu \Gamma (2 \nu -1/2)}{\Gamma (2\nu +1)} \sum_{\ell =1}^\infty \frac{[\sin\rho]^{2\ell}}{\ell^{(4\nu-1)}} 
    = \frac{2^{4\nu }\sqrt{\pi}\nu \Gamma (2 \nu -1/2)}{\Gamma (2\nu +1)} {\text{Li}}_{(4\nu-1)}(\sin^2\rho ) ,
    \label{eq:VPB7}
\end{equation}
where $\text{Li}_{4 \nu-1}[\sin^2\rho]$ is the polylogarithm function \cite[25.12.10]{NIST:DLMF}.
To find the behaviour of \eqref{eq:VPB7}
as $\rho \to \pi/2$, we 
use the relationship between the polylogarithm and Lerch's transcendental function  $\Phi _{L}$
\cite[25.14.1]{NIST:DLMF}
\begin{equation}
     {\text{Li}}_{(4\nu-1)}(\sin ^{2}\rho ) = (\sin ^{2}\rho ) \Phi _{L}(\sin ^{2}\rho , 4\nu - 1, 1) ,
\end{equation}
together with the result in \cite[64:12.7]{atlas}:
\begin{equation}
    \lim _{\rho \to \pi /2} \left[ \frac{\Phi _{L}(\sin ^{2}\rho,4\nu -1,1)}{(\cos ^{2}\rho )^{4\nu - 2}}\right] = \Gamma (2-4\nu ).
    \label{eq:Lerch limit}
\end{equation} 
However, \eqref{eq:Lerch limit} is only valid for $\nu <1/2$. So, for $1/4<\nu <1/2$, we have, as $\rho \to \pi /2$,
\begin{equation}
    \Sigma  _{1}\sim \frac{2^{4\nu }\sqrt{\pi}\nu \Gamma (2 \nu -1/2)\Gamma (2-4\nu )}{\Gamma (2\nu +1)} [\cos \rho ] ^{8\nu - 4}. 
    \label{eq:VPB8}
\end{equation}
Since the divergences arise from the large $\lvert \ell \rvert $, $\lvert \omega \rvert $ behaviour which is captured in $\Sigma _{1}$  (and thence in $\Sigma _{0}$), the dominant behaviour of $\langle {\hat {\Phi }}^{2} \rangle _{0}^{\zeta }- \langle {\hat {\Phi }}^{2} \rangle _{0}^{N}$ at the space-time boundary is obtained by substituting $4\Sigma _{1}$ for the sum and integral (to account for both positive and negative values of $\omega $ and $\ell $) in \eqref{eq:VPB4} and subsequently in \eqref{eq:VPB3}, which gives
\begin{equation}
    \langle {\hat {\Phi }}^{2} \rangle _{0}^{\zeta } -\langle {\hat {\Phi }}^{2} \rangle _{0}^{N}
    \sim 
    ~ -\frac{2^{4\nu }\Gamma (\nu)^2\Gamma(2\nu -1/2)\Gamma (2-4\nu )\cot \zeta }{2 \pi^{3/2}\,L\, \Gamma(-\nu)^2 \Gamma(2\nu+1)} [\cos \rho]^{6\nu-2},
\end{equation}
which vanishes in the limit $\rho \to \pi/2$ when $\nu >1/3$.
Therefore for $1/3<\nu<1/2$ we have 
$\langle {\hat {\Phi }}^{2} \rangle _{0}^{\zeta }= \langle {\hat {\Phi }}^{2} \rangle _{0}^{N}$ at the space-time boundary. 

When $\nu>1/2$ we note that polylogarithm function in \eqref{eq:VPB7} becomes the  Riemann zeta function as $\rho\to \pi/2$ \cite[25.12.10]{NIST:DLMF}, which is finite for these values of $\nu $. When we then substitute \eqref{eq:VPB7} back into \eqref{eq:VPB3} the right-hand-side vanishes due to the $[\cos \rho] ^{2-2\nu }$ term since $\nu <1$. Therefore, 
from our analysis of the quantity $\Sigma _{1}$, we deduce that $\langle {\hat {\Phi }}^{2} \rangle _{0}^{\zeta }= \langle {\hat {\Phi }}^{2} \rangle _{0}^{N}$ at the space-time boundary for $\nu >1/3$, as the case $\nu=1/2$ was considered earlier.

While the above discussion is for v.e.v.s, similar considerations apply to t.e.v.s.
In that case the integral over $\omega $ in \eqref{eq:VPB4} is replaced by a sum over $n$ (from replacing $\omega $ by $n\kappa $), leading us to consider the quantity
\begin{equation}
    \Sigma _\beta = \sum _{\ell =1}^{\infty } \sum _{n=1}^{\infty } \frac{[\sin\rho] ^{2\ell}}{(\ell^2 + n^2 \kappa ^2)^{2\nu}} .
\end{equation}
While the sum is convergent for $\nu > 1/4$, it cannot be performed analytically. However, using the Euler-Maclaurin formula \cite[2.10.1]{NIST:DLMF}, we have the asymptotic approximation
\begin{equation}
    \sum _{n=0}^{\infty } \frac{1}{(\ell^2 + n^2 \kappa ^2)^{2\nu}}  \sim 
    \int _{0}^{\infty } \frac{1}{(\ell^2 + y^2 \kappa ^2)^{2\nu}} dy + \frac{1}{\ell ^{4\nu }} + \ldots ,
    \label{eq:VPB11}
    \end{equation}
    where $\ldots $ denotes terms which vanish more rapidly as $\ell \rightarrow \infty $.
    Therefore the dominant behaviour in $\Sigma _{\beta }$, for large $\ell $, arises from the integral in \eqref{eq:VPB11}. 
    The same analysis as that performed for v.e.v.s then leads us to deduce that $\langle {\hat {\Phi }}^{2} \rangle _{\beta }^{\zeta }= \langle {\hat {\Phi }}^{2} \rangle _{\beta }^{N}$ at the space-time boundary for $\nu >1/3$.

The above analytic argument is valid only for $\nu > 1/3$. 
However, our numerical results in section~\ref{sec:VP_Robin} indicate that $\langle {\hat {\Phi }}^{2} \rangle ^{\zeta }= \langle {\hat {\Phi }}^{2} \rangle ^{N}$ on the space-time boundary for both v.e.v.s and t.e.v.s and values of $\nu $ less than or equal to $1/3$.
In the absence of an analytic argument, we now present some additional numerical evidence for v.e.v.s (similar results are obtained for t.e.v.s).  
Using the form \eqref{eq:g1} for the radial function $p_{\omega \ell }(\rho )$, and applying \cite[15.8.1]{NIST:DLMF}, we consider the quantity
\begin{equation}
   \Sigma _{2} = \sum_{\ell=-\infty}^{\infty} \int^{\infty}_{-\infty} d \omega\, \mathcal{N}_{\omega \ell}^\zeta \,\, \mathcal{N}_{\omega \ell}^N \, (\sin\rho)^{2\lvert \ell \rvert } 
   [F ( b-c+1, a-c+1, a + b - c +1 ; \sin ^{2}\rho ) ]^{2},
   \label{eq:nocosVPB2}
\end{equation}
where the parameters $a$, $b$ and $c$ are given in \eqref{eq:alphabetaEuc}.
The quantity $\Sigma _{2}$ will be multiplied by a factor $[\cos \rho ]^{2-2\nu }$ (plus some numerical factors) to give $\langle {\hat {\Phi }}^{2} \rangle ^{\zeta }- \langle {\hat {\Phi }}^{2} \rangle ^{N}$.
Therefore, if we can demonstrate that $\Sigma _{2}$ is finite as the boundary is approached, it must be the case that $\langle {\hat {\Phi }}^{2} \rangle ^{\zeta }= \langle {\hat {\Phi }}^{2} \rangle ^{N}$ on the boundary, as required.
The hypergeometric function in \eqref{eq:nocosVPB2} is regular as $\rho \to \pi /2$ and the boundary is approached. However, setting $\rho \to \pi /2$ in \eqref{eq:nocosVPB2} (and replacing the numerical factors and appropriate powers of $\cos \rho $) simply gives \eqref{eq:VPB3}, which we have already examined.  
We therefore explore the behaviour of $\Sigma _{2}$ for $\rho < \pi /2$, as follows. 

\begin{figure}[htbp]
     \begin{center}
        \includegraphics[width=0.8\textwidth]{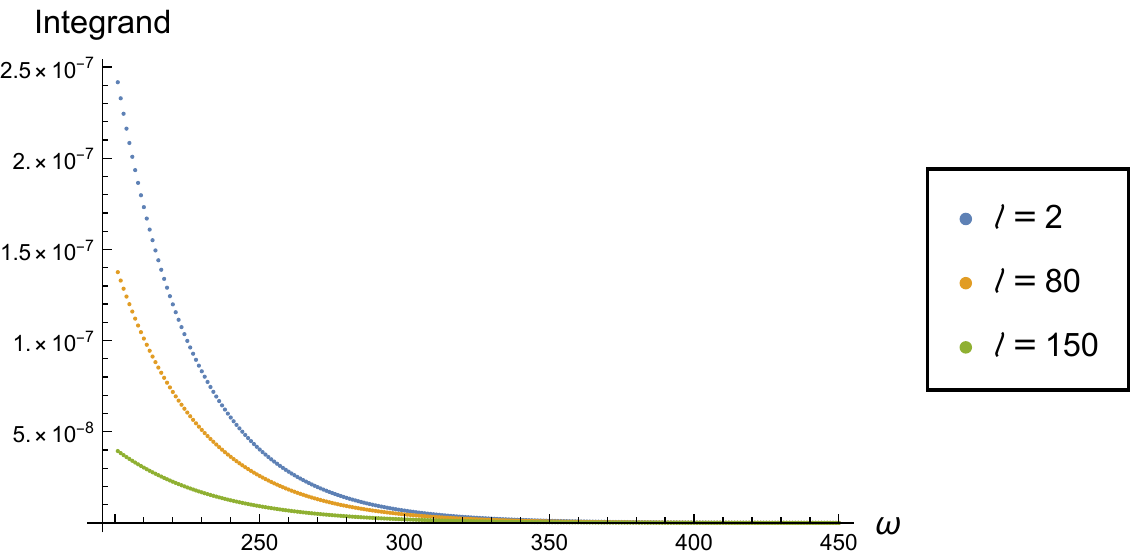} 
               \caption{The integrand in \eqref{eq:nocosVPB2} for $\rho=99\pi/100$, $\nu=1/8$, $\zeta=\pi/3$, $\omega \in [201,450]$ and a selection of values of $\ell $.}
        \label{fig:w_200}
      \end{center}
\end{figure}
\begin{figure}[htbp]
     \begin{center}
        \includegraphics[width=0.8\textwidth]{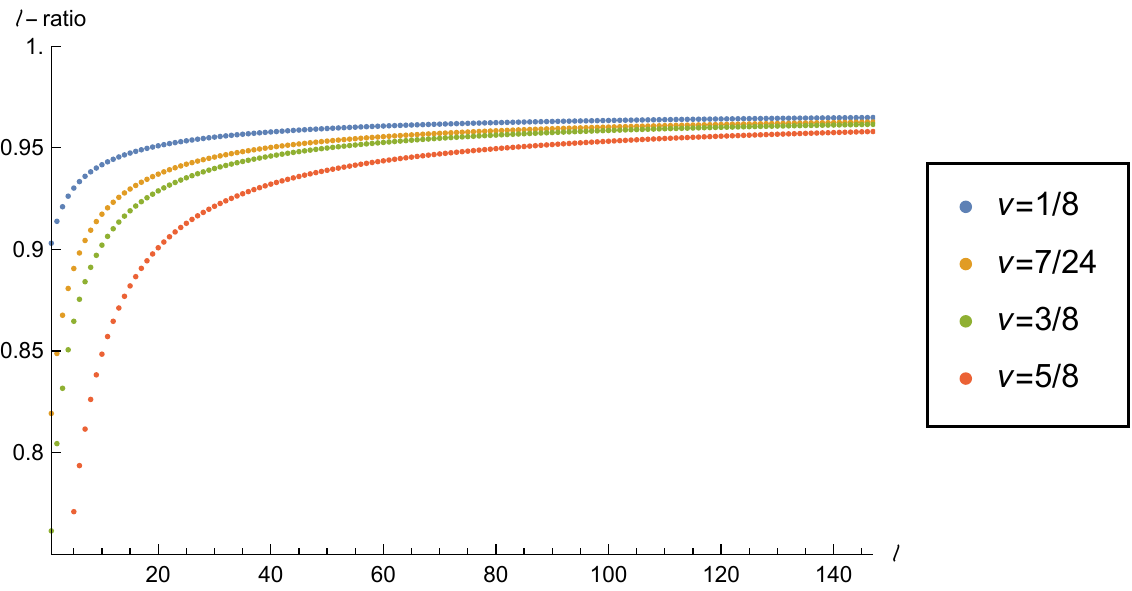} 
               \caption{Ratio of successive terms in the sum over $\ell $ in \eqref{eq:nocosVPB2} for four different values of $\nu$ with $\rho=99\pi/200$ and $\zeta =\pi/3$. In each case the integral over $\omega $ has been performed for $\lvert \omega \rvert \le 200$.}
        \label{fig:l_summand1}
      \end{center}
\end{figure}

For fixed $\ell $ and $\rho < \pi /2$, the integral over $\omega $ in \eqref{eq:nocosVPB2} converges very rapidly for large $\omega $, as can be seen in Figure~\ref{fig:w_200}. 
We compute the integral over $\omega $ numerically  for fixed $\ell $, for $\lvert \omega \rvert \le 200$.
The integral for larger values of $\lvert \omega \rvert $ is estimated by fitting a sum of exponentials to the integrand in the region $\omega \in [201,700]$ and then integrating the fitting functions. We thereby estimate that the relative error in truncating the integral at $\lvert \omega \rvert =200$ is of the order of $10^{-4}$.

We then use the ratio test to show that the  sum over $\ell$ converges. Our results for $\rho = 99\pi /200$ and four different values of  $\nu$ are shown in figure \ref{fig:l_summand1}. We plot the ratio of successive terms in the sum over $\ell$ for $\ell \le 150$. For all values of $\nu$, the ratio of successive terms in the sum over $\ell$, as $\ell $ increases, tends to a limit which is below unity. 
Our numerical investigations reveal that the limit is below unity for all $0<\rho < \pi /2$, but that the limit increases towards unity as $\rho \to \pi /2$. This provides evidence that the sum over $\ell$ converges to a finite value for $0<\rho < \pi/2$. 

To find $\langle {\hat {\Phi }}^{2} \rangle _{0}^{\zeta }- \langle {\hat {\Phi }}^{2} \rangle _{0}^{N}$ at the space-time boundary,  we multiply the finite value of $\Sigma _{2}$ obtained from (\ref{eq:nocosVPB2}) by $[\cos\rho]^{2-2\nu }$, which vanishes as $\rho \to \pi /2$ (there are also some irrelevant numerical factors). 
We conclude $\langle {\hat {\Phi }}^{2} \rangle _{0}^{\zeta }=\langle {\hat {\Phi }}^{2} \rangle _{0}^{N}$
at the space-time boundary for all values of $\nu \in (0,1)$.

\section{Conclusions} 
\label{sec:conc}
 In this paper we have studied the renormalised v.e.v.s and t.e.v.s of the VP of a real quantum scalar field, with general mass and coupling, propagating on three-dimensional adS space-time. 
 We have constructed the Feynman's Green's functions for both the vacuum and thermal states and have used Hadamard renormalisation to calculate the v.e.v.s and t.e.v.s of the VP, applying  Dirichlet and Neumann boundary conditions at the time-like boundary. 
The v.e.v.s with both Dirichlet and Neumann boundary conditions respect the maximum symmetry of the background adS space-time, whereas this symmetry is lost for the thermal states,  resulting in t.e.v.s which depend on the space-time position $\rho $. As the space-time boundary is approached ($\rho \to \pi/2$), the t.e.v.s with both Dirichlet and Neumann boundary conditions reach their respective v.e.v.s for all values of the inverse temperature $\beta$.

We have also determined the v.e.v.s and t.e.v.s of the VP with Robin boundary conditions applied at the time-like boundary. This was achieved by studying the scalar field in Euclidean space as the Euclidean Green's function is uniquely prescribed. Our findings show that the  v.e.v.s and t.e.v.s for all $\nu$ and  Robin parameter $\zeta$ approach the  Neumann values at the space-time boundary  except for  Dirichlet which has a different limit. 

Earlier work determined  the VP on $n$-dimensional  adS \cite{Kent:2014nya} for a scalar field with general mass and coupling but only with Dirichlet boundary conditions, whilst~\cite{Morley:2021}  looked at Dirichlet, Neumann and Robin boundary conditions for a scalar field but only in the massless conformally coupled case in four-dimensional adS. We have extended the work in these papers by considering the VP for a scalar field with general mass and coupling with Dirichlet, Neumann and Robin boundary conditions. 

Our result that the v.e.v.s and t.e.v.s with Robin boundary conditions converge to the Neumann case at the space-time boundary matches the finding in~\cite{Morley:2021} for the massless conformally coupled case. We conclude that this result is a general property of quantum scalar field theory on adS, namely that the generic behaviour of the fields at the space-time boundary is given by Neumann boundary conditions, while Dirichlet boundary conditions, although they are the most widely considered in the literature, give expectation values with different behaviour at the boundary. 

We may understand this heuristically as follows.  
For Dirichlet boundary conditions, the scalar field decays as rapidly as possible at the space-time boundary.  For all other boundary conditions, the leading-order behaviour of the scalar field as the boundary is approached is the same. 
Furthermore, as observed in \cite{Dappiaggi:2018xvw}, while Dirichlet boundary conditions uniquely specify the form of the scalar field near the boundary, the notion of Neumann boundary conditions is not, in general, unique. 
We have made a particular choice of boundary conditions to be defined as Neumann boundary conditions.
A different choice of Neumann boundary conditions would correspond to one of our Robin boundary conditions. 
Since the value of the VP on the boundary is the same for all boundary conditions other than Dirichlet, our results demonstrate that all possible choices of Neumann boundary conditions yield the same value of the VP on the boundary.
The differences between Neumann and Robin boundary conditions appear only at subleading order and hence do not affect the v.e.v.s and t.e.v.s of the scalar field on the boundary.

In this work we have considered the VP of a quantum scalar field. 
It would be very interesting to study the renormalized SET, which governs the back-reaction of the quantum scalar field on the space-time geometry.  
The v.e.v.~of the SET for a scalar field with arbitrary mass and coupling, and Dirichlet boundary conditions applied, leads to a renormalization of the cosmological constant in the semi-classical Einstein equations in any number of dimensions \cite{Kent:2014nya}. 
We expect that a similar result holds for Neumann boundary conditions, since the vacuum state in this case is also maximally symmetric, although the explicit computation of the SET for these boundary conditions has yet to be completed.  
T.e.v.s for all boundary conditions, and v.e.v.s for Robin boundary conditions, are no longer maximally symmetric so the solution of the back-reaction problem in these states is likely to be more complicated. 

Finally, this paper contains preliminary work for a study of quantum scalar field theory on a BTZ black hole \cite{Banados:1992gq, Banados:1992wn}.
Since the BTZ metric is constructed by identifying points in three-dimensional adS \cite{Banados:1992gq, Banados:1992wn}, the Green's function for a massless, conformally coupled scalar field on BTZ can be found using the method of images \cite{Steif:1993zv, Lifschytz:1993eb}.
This greatly facilitates the calculation of the renormalized VP and SET, which have been computed for a massless, conformally coupled scalar field with transparent \cite{Steif:1993zv},
Dirichlet \cite{Lifschytz:1993eb, Kothawala:2008sm} and Neumann \cite{Lifschytz:1993eb} boundary conditions (see also \cite{Shiraishi:1993nu, Shiraishi:1993qnr}). We do not consider the former in our work here since they can be applied only to a massless and conformally coupled scalar field. 
The corresponding calculation for a rotating BTZ black hole with Robin boundary conditions is likely to be complicated by the fact that, while there are no superradiant modes when either Dirichlet or Neumann boundary conditions are applied \cite{Ortiz:2011wd}, superradiant modes exist for a set of values of the Robin parameter $\zeta $ \cite{Dappiaggi:2017pbe}. 
Green's functions for the ground state can be constructed for Robin boundary conditions \cite{Bussola:2017wki}, but the VP and SET remain to be computed.  
We plan to explore this situation further in future work.

\section*{Declarations}
 The authors declare that they have no conflict of interest.
The work of E.W.~is supported by the Lancaster-Manchester-Sheffield Consortium for Fundamental Physics under STFC grant ST/T001038/1.
This research has also received funding from the European Union's Horizon 2020 research and innovation program under the H2020-MSCA-RISE-2017 Grant No.~FunFiCO-777740.
Data supporting this publication can be freely downloaded from the University of Sheffield Research Data Repository at {\url {https://doi.org/10.15131/shef.data.21221009}}, under the terms of the Creative Commons Attribution (CC BY) licence.
The version of record of this article, first published in {\it {General Relativity and Gravitation}}, is available online at the Publisher’s website: {\url {https://dx.doi.org/10.1007/s10714-022-03056-6}}.

\bibliographystyle{spphys}       

\end{document}